\documentclass{article}

\usepackage{arxiv}

\usepackage[utf8]{inputenc} % allow utf-8 input
\usepackage[T1]{fontenc}    % use 8-bit T1 fonts
\usepackage{url}            % simple URL typesetting
\usepackage{booktabs}       % professional-quality tables
\usepackage{amsfonts}       % blackboard math symbols
\usepackage{nicefrac}       % compact symbols for 1/2, etc.
\usepackage{microtype}      % microtypography
\usepackage{lipsum}
\usepackage{graphicx}
\usepackage{orcidlink}
\usepackage{comment}

\usepackage[numbers]{natbib}
\usepackage{longtable}
\usepackage{caption}
\usepackage{amsmath}

\definecolor{VUB_blauw}{rgb}{0.1529, 0.2667, 0.5529}
\usepackage{hyperref}       % hyperlinks
\hypersetup{
    colorlinks,%
    citecolor=VUB_blauw,%
    filecolor=VUB_blauw,%
    linkcolor=VUB_blauw,%
    urlcolor=VUB_blauw
}
\graphicspath{ {./images/} }

\AddToHook{shipout/background}{%
  \ifnum\value{page}=1 % Check if the current page is 2
    \pagenumbering{Roman} % Change numbering to Roman numerals
    \setcounter{page}{1} % Set the page number to II
  \fi
  \ifnum\value{page}=2 % Check if the current page is 2
  \pagenumbering{arabic} % Revert to Arabic numerals
  \setcounter{page}{2} % Set the page number to 3
  \fi
}

\title{How Deep Do Large Language Models Internalize Scientific Literature and Citation Practices?}
\runningtitle{How Deep Do Large Language Models Internalize Scientific Literature and Citation Practices?}

\author{
  Andres Algaba\textsuperscript{1,$\ast$} \\
  \orcidlinkc{0000-0002-0532-3066} \\
  \And
  Vincent Holst\textsuperscript{1} \\ 
  \orcidlinkc{0009-0002-4117-4966} \\
  \And
  Floriano Tori\textsuperscript{1} \\ 
  \orcidlinkc{0000-0003-4358-3368} \\
  \And
  Melika Mobini\textsuperscript{1} \\ 
  \orcidlinkc{0009-0001-0580-1072} \\
  \And
  Brecht Verbeken\textsuperscript{1} \\ 
  \orcidlinkc{0000-0002-7506-3298} \\
  \And
  Sylvia Wenmackers\textsuperscript{2} \\ 
  \orcidlinkc{0000-0002-1041-3533} \\
  \And
  Vincent Ginis\textsuperscript{1,3} \\ 
  \orcidlinkc{0000-0003-0063-9608} \\
  \and
  \textsuperscript{1}Data Analytics Lab, Vrije Universiteit Brussel, 1050 Brussel, Belgium \\ 
  \textsuperscript{2}Centre for Logic and Philosophy of Science (CLPS), KU Leuven, 3000 Leuven, Belgium \\ 
  \textsuperscript{3}School of Engineering and Applied Sciences, Harvard University, Cambridge, Massachusetts 02138, USA
}

\begin{document}

\maketitle

\renewcommand{\thefootnote}{}
\footnotetext{$\ast$ Corresponding author: \href{mailto:andres.algaba@vub.be}{andres.algaba@vub.be} \\ Code available at: \url{https://github.com/AndresAlgaba/LLMs_scientific_literature} \\ Data available at: \url{https://zenodo.org/record/15124610}}

\renewcommand{\thefootnote}{\arabic{footnote}}

\thispagestyle{plain} 

\begin{abstract}
The spread of scientific knowledge depends on how researchers discover and cite previous work. The adoption of large language models (LLMs) in the scientific research process introduces a new layer to these citation practices. However, it remains unclear to what extent LLMs align with human citation practices, how they perform across domains, and may influence citation dynamics. Here, we show that LLMs systematically reinforce the Matthew effect in citations by consistently favoring highly cited papers when generating references. This pattern persists across scientific domains despite significant field-specific variations in existence rates, which refer to the proportion of generated references that match existing records in external bibliometric databases. Analyzing $274,951$ references generated by GPT-4o for $10,000$ papers, we find that LLM recommendations diverge from traditional citation patterns by preferring more recent references with shorter titles and fewer authors. Emphasizing their content-level relevance, the generated references are semantically aligned with the content of each paper at levels comparable to the ground truth references and display similar network effects while reducing author self-citations. These findings illustrate how LLMs may reshape citation practices and influence the trajectory of scientific discovery by reflecting and amplifying established trends. As LLMs become more integrated into the scientific research process, it is important to understand their role in shaping how scientific communities discover and build upon prior work.
\end{abstract}

% keywords can be removed
% \keywords{Automated Science \and Large Language Models \and Responsible AI Practices \and Science of Science}

\begin{figure}[t!]
	\centering
	\includegraphics[width=1\textwidth]{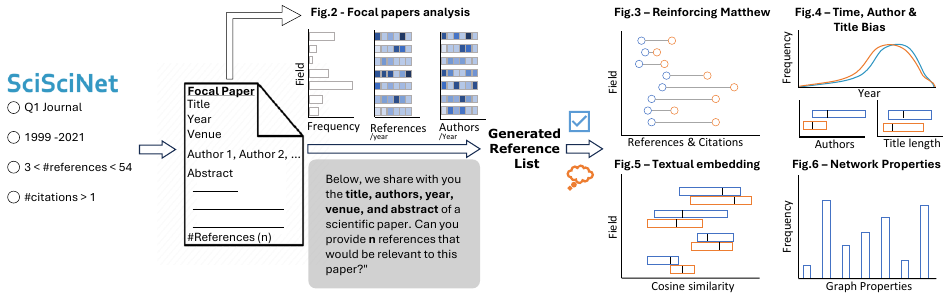}
	\caption{\textbf{Overview of our experiment comparing the characteristics of human citations and LLM generated references, when tasked to suggest references based on the title, authors, year, venue, and abstract of a paper.} We sample $10,000$ focal papers from all SciSciNet \citep{lin2023sciscinet} papers which are published in Q1 journals between $1999$ and $2021$, have in between 3 and 54 references, and have at least 1 or more citations (n=$17,538,900$). We prompt GPT-4o to generate suggestions of references based on the title, authors, year, venue, and abstract of a focal paper, where the number of requested generated references corresponds to the ground truth number of references made in the focal paper, which amounts to a total of $274,951$ references. We verify the existence of the generated references via the SciSciNet \citep{lin2023sciscinet} database and compare the characteristics, such as title length, publication year, venue, number of authors, and semantic embeddings, of the existing and non-existent generated references with the ground truth. For the existing generated references, we also compare additional characteristics, such as the number of citations and references, and analyze the properties of their citation networks.}
	\label{fig:main_1}
\end{figure}

\section*{Introduction}
Large Language Models (LLMs) have evolved beyond simple natural language processing tasks, demonstrating remarkable capabilities in, for example, mathematical problem-solving and code generation \citep{brown2020language,bubeck2023sparks}. This expansion of capabilities has led to rapid adoption in scientific research \citep{bick2024rapid,mishra2024use}, where LLMs are now being applied to support data science analysis~\citep{hollmann2023automated,manning2024automatedsocialsciencelanguage}, assist in experiments~\citep{han2024mining,hewitt2024predictions,lehr2024chatgpt,ziems2024can}, and integrate into larger systems for scientific discovery~\citep{Bago2024science,m2024augmenting,merchant2023scaling,romera2024mathematical,trinh2024solving,zheng2025large}. Since scientists are even experimenting with LLMs as autonomous research agents~\citep{baek2024researchagent,boiko2023emergent,ghafarollahi2024sciagents,gottweis2025towards,lu2024aiscientist,schmidgall2025agentlaboratoryusingllm,si2024can,swanson2024virtual}, we focus on their potential to assist in scientific synthesis \citep{delgado2025review,dennstadt2024title,gougherty2024testing,kang2024researcharena,skarlinski2024language,susnjak2024automating} -- a development that could significantly impact the dissemination of scientific information and potentially skew the landscape of scientific knowledge \citep{greenberg2009citation,schneider2024something,simkin2002read,tahamtan2018core}.

Current approaches to LLM-assisted scientific synthesis predominantly use document retrieval methods \citep{gao2023retrieval,lewis2020retrieval}, complemented by model fine-tuning \citep{li2024scilitllm,susnjak2024automating} or citation network navigation \citep{skarlinski2024language}. However, the reliance on search methods within external databases is unlikely to introduce significant new information dissemination patterns, potentially leading to effects similar to those observed with the introduction of comprehensive academic search tools \citep{evans2008electronic,fortunato2018science,nielsen2021global}. We argue that as LLMs improve their capabilities of generating reference suggestions through the parametric knowledge acquired during training \citep{algaba2024largelanguagemodelsreflect}, scientists could leverage their internal world models \citep{gurnee2024languagemodelsrepresentspace,hao2023reasoning} to make more informed decisions about which prior work to cite and build upon \citep{Trujillo2018cocitations}. However, it remains unclear to what extent LLMs align with human citation practices, perform across domains, and may influence existing citation patterns.

We address these questions on LLM-generated recommendations by comparing the characteristics of human citation patterns and reference suggestions generated by LLMs solely relying on their parametric knowledge. Scientific literature has long exhibited well-documented characteristics and potential biases in citation practices \citep{bornmann2008citation,liu2023review,smith2012impact}, which vary across fields \citep{radicchi2008universality,wang2013quantifying,Fang2024disciplines}, including preferences for recent publications \citep{tahamtan2016factors}, articles with shorter titles \citep{letchford2015advantage}, papers from high-profile venues \citep{lawrence2003politics}, and works with multiple authors \citep{gazni2011investigating,thelwall2023coauthored}, as well as the ``Matthew effect,'' by which highly cited papers accumulate even more citations over time \citep{mattheweffect,price1965networks,wang2014unpacking}. We argue that LLMs may introduce or exacerbate (field-specific) citation patterns \citep{acerbi2023large,checco2021ai,horbach2022automated,navigli2023biases,tjuatja2024llms} due to their struggle with long-tail knowledge \citep{kandpal2023longtail} combined with uneven representation of scientific work in LLMs' training data \citep{dubey2024llama,team2023gemini}. While our experimental setup may not fully reflect real-world usage of LLMs for citation generation, which often involves more interactivity and reliance on external data sources, it provides a controlled laboratory setting to assess the parametric knowledge and inherent citation patterns of LLMs \citep{bubeck2023sparks}. Understanding these patterns is crucial as they may influence how researchers discover and cite prior work, potentially reinforcing existing biases in scientific literature and shifting citation practices over time \citep{greenberg2009citation,schneider2024something,simkin2002read,tahamtan2018core}.

\begin{figure}[t!]
	\centering
	\includegraphics[width=1.0\textwidth]{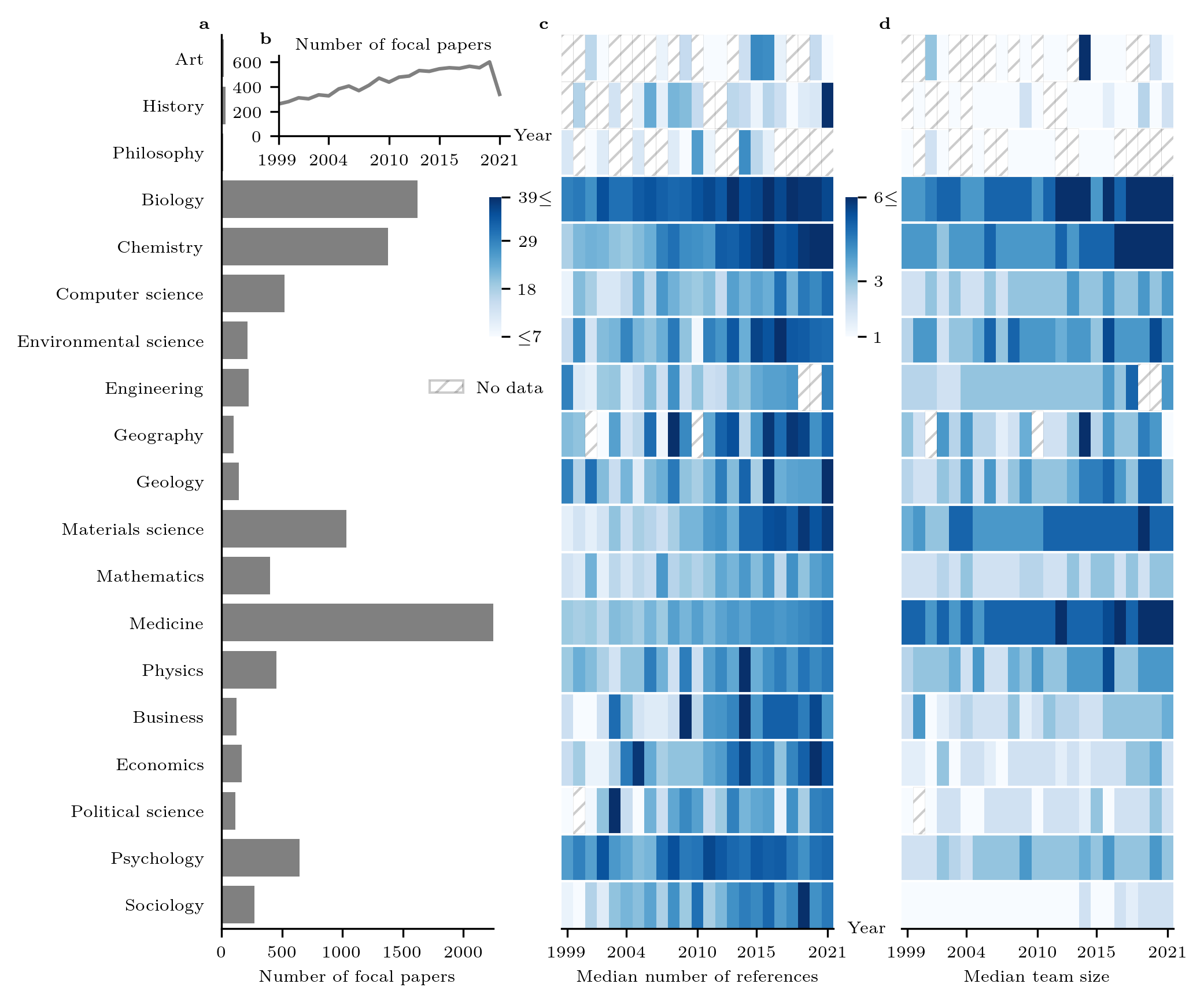}
	\caption{\textbf{Descriptive statistics of the focal paper sample.} This figure summarizes key characteristics of the focal paper sample (n=$10,000$) across fields. \textbf{a}, The distribution of focal papers by field highlights strong representation in the exact sciences (biology, chemistry, computer science, environmental science, engineering, geography, geology, materials science, mathematics, medicine, and physics), with comparatively fewer papers in the humanities (art, history, and philosophy) and the social sciences (business, economics, political science, psychology, and sociology) (Appendix Table \ref{tab:mapping}). \textbf{b}, The temporal trend in the number of focal papers exhibits linear growth from 1999 until 2021, which aligns with full SciSciNet \citep{lin2023sciscinet} database for this period. \textbf{c},\textbf{d}, Both the median number of references cited per focal paper and the median team size are increasing over time. This pattern is more clear in the fields with a larger number of focal papers (e.g., biology, chemistry, and medicine). The color intensity represents the magnitude of the values: darker shades indicate higher numbers, while lighter shades represent lower values. Hatched cells indicate no data available for a given year and field.}
	\label{fig:main_2}
\end{figure}

\section*{Experimental setup for generating references}
Figure \ref{fig:main_1} displays our experimental setup where we sample $10,000$ focal papers from the SciSciNet \citep{lin2023sciscinet} database which are published in Q1 journals between $1999$ and $2021$, have in between 3 and 54 references, and have at least 1 or more citations. Additionally, we require the sampled focal paper to have a ``top field'' and a valid DOI with an abstract (n=$17,538,900$, see \hyperref[sec:appendix]{Appendix}). We then prompt GPT-4o to generate suggestions of references based on the title, authors, year, venue, and abstract of a focal paper, where the number of requested generated references corresponds to the ground truth number of references made in the focal paper, which amounts to a total of $274,951$ references.

LLMs are known to produce hallucinations or confabulations -- generating content that contains factual errors or makes claims unsupported by specific sources, such as incorrect statements about historical events \citep{huang2023survey}. Since LLMs only provide suggestions, researchers can verify both existence and appropriateness through external databases \citep{agrawal-etal-2024-language,fabiano2024optimize,simkin2002read}. However due to automation bias~\citep{skitka1999does}, these suggestions may nonetheless steer their search process for discovering and building upon prior work~\citep{greenberg2009citation}. In our experiment, we cross-check all generated references against the SciSciNet \citep{lin2023sciscinet} database using fuzzy matching, classifying references as ``existing'' based on conservative similarity thresholds, likely underestimating the true existence rate due to these methodological choices (see \hyperref[sec:appendix]{Appendix}).

Our experimental setup differs from both traditional LLM benchmarks \citep{chen2021evaluating,clark2018think,cobbe2021training,hendrycks2020measuring,hendrycks2021measuring,jimenez2024swebench,phan2025humanity,rein2023gpqa,srivastava2022beyond} and previous assessments of LLMs in scientific literature contexts \citep{agrawal-etal-2024-language,ajith2024litsearch,algaba2024largelanguagemodelsreflect,chen2025unveiling} in several key aspects. Rather than asking LLMs to attribute open-ended scientific claims \citep{press2024citeme}, write literature reviews \citep{kang2024researcharena,walters2023fabrication}, or suggest in-text references \citep{algaba2024largelanguagemodelsreflect}, we provide paper abstracts as input. Abstracts contain more condensed information, reducing reliance on memorization~\citep{Chen_Lin_Han_Sun_2024,kadavath2022language}, while their standardized format across scientific fields and concise nature better reflect how researchers might query these models in practice. Moreover, while previous work primarily evaluates suggested references based on their existence, bibliometric accuracy, or expert judgment \citep{qureshi2023chatgpt}, we analyze how LLMs' citation patterns align with human citation practices. Finally, we focus on the LLM's parametric knowledge acquired during training, without enhancing its capabilities through search or retrieval-augmented generation \citep{kang2024researcharena,lewis2020retrieval,m2024augmenting}. This focus on parametric knowledge allows us to examine LLMs as reasoning engines that may reshape how scientists discover connections in literature \citep{gurnee2024languagemodelsrepresentspace,Truhn2023reasoning}, distinct from search-based approaches that may largely reflect existing information dissemination patterns \citep{evans2008electronic}.

Figure~\ref{fig:main_2} summarizes key characteristics of the focal paper sample (n=$10,000$) across fields. The distribution of papers shows a clear predominance of exact sciences, particularly medicine, biology, and chemistry, with substantially fewer papers from humanities and social sciences (Figure~\ref{fig:main_2}\textbf{a}), reflecting the relative field proportions in the full SciSciNet \citep{lin2023sciscinet} database. The temporal trend in the number of focal papers exhibits linear growth from 1999 until 2021, which aligns with full SciSciNet \citep{lin2023sciscinet} database for this period. The commonly reported exponential growth in scientific publications is observed in the decades preceding this period. We also observe two distinctive temporal trends consistent with broader patterns in scientific publishing: a steady increase in both the median number of references per paper (Figure~\ref{fig:main_2}\textbf{c}) and median team size (Figure~\ref{fig:main_2}\textbf{d}). These trends compare to previous findings on the expansion of reference lists \citep{holst2024dataset} and the growing prevalence of larger research teams \citep{lin2023sciscinet,wu2019large}, suggesting our sample captures key dynamics in modern scientific practice.

\begin{figure}
	\centering
	\includegraphics[width=1\textwidth]{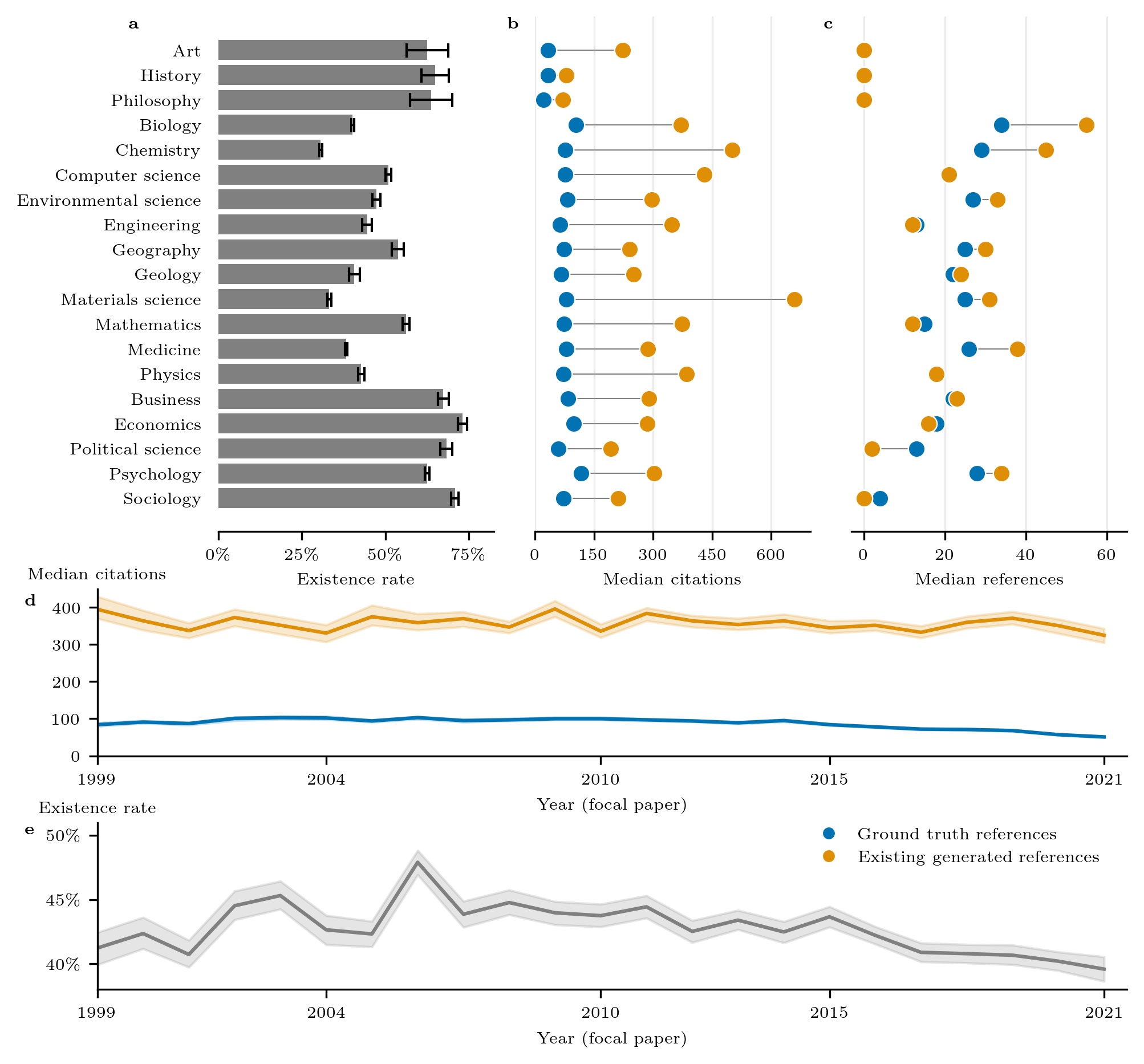}
	\caption{\textbf{Existing generated references reinforce the Matthew effect in citations.} This figure displays the existence rate of generated references (gray, n=$274,497$), and the citation characteristics of the ground truth (blue, n=$274,951$) and existing generated (orange, n=$116,939$) references across fields and time. Error bars and shaded bands represent $95\%$ confidence intervals. \textbf{a}, The existence rate of generated references by field of the focal paper shows significantly lower values in the exact sciences compared to the humanities and the social sciences. \textbf{b}, Median citation counts reveal that existing generated references tend to have higher citation counts across all fields, suggesting a preference toward already highly cited works. The pairwise two-sided Wilcoxon signed-rank test at the focal paper level confirms that the existing generated references have a statistically significant higher median citation count for all fields (history, $p$=$0.003$; philosophy, $p$=$0.022$: all other fields, $p$$<$$0.001$). \textbf{c}, The median reference counts tend to be more similar for many fields, with only political science showing existing generated references to have a lower median number reference count. The pairwise two-sided Wilcoxon signed-rank test at the focal paper level shows that the existing generated references have a statistically significant higher median reference count for biology ($p$$<$$0.001$), chemistry ($p$$<$$0.001$), environmental science ($p$$<$$0.001$), geography ($p$=$0.007$), materials science ($p$$<$$0.001$), mathematics ($p$$<$$0.001$), medicine ($p$$<$$0.001$), psychology ($p$$<$$0.001$), and sociology ($p$=$0.002$). All other fields show no statistically significant difference ($p$$>$$0.05$). \textbf{d}, Temporal trends at the focal paper level show that existing generated references consistently exhibit higher median citation counts compared to ground truth references, further emphasizing the reinforcement of the Matthew Effect in citations. \textbf{e}, The overall existence rate of generated references remains consistent across the publication year of the focal paper, fluctuating between $40\%$ and $50\%$.}
	\label{fig:main_3}
\end{figure}

\begin{figure}
	\centering
	\includegraphics[width=1\textwidth]{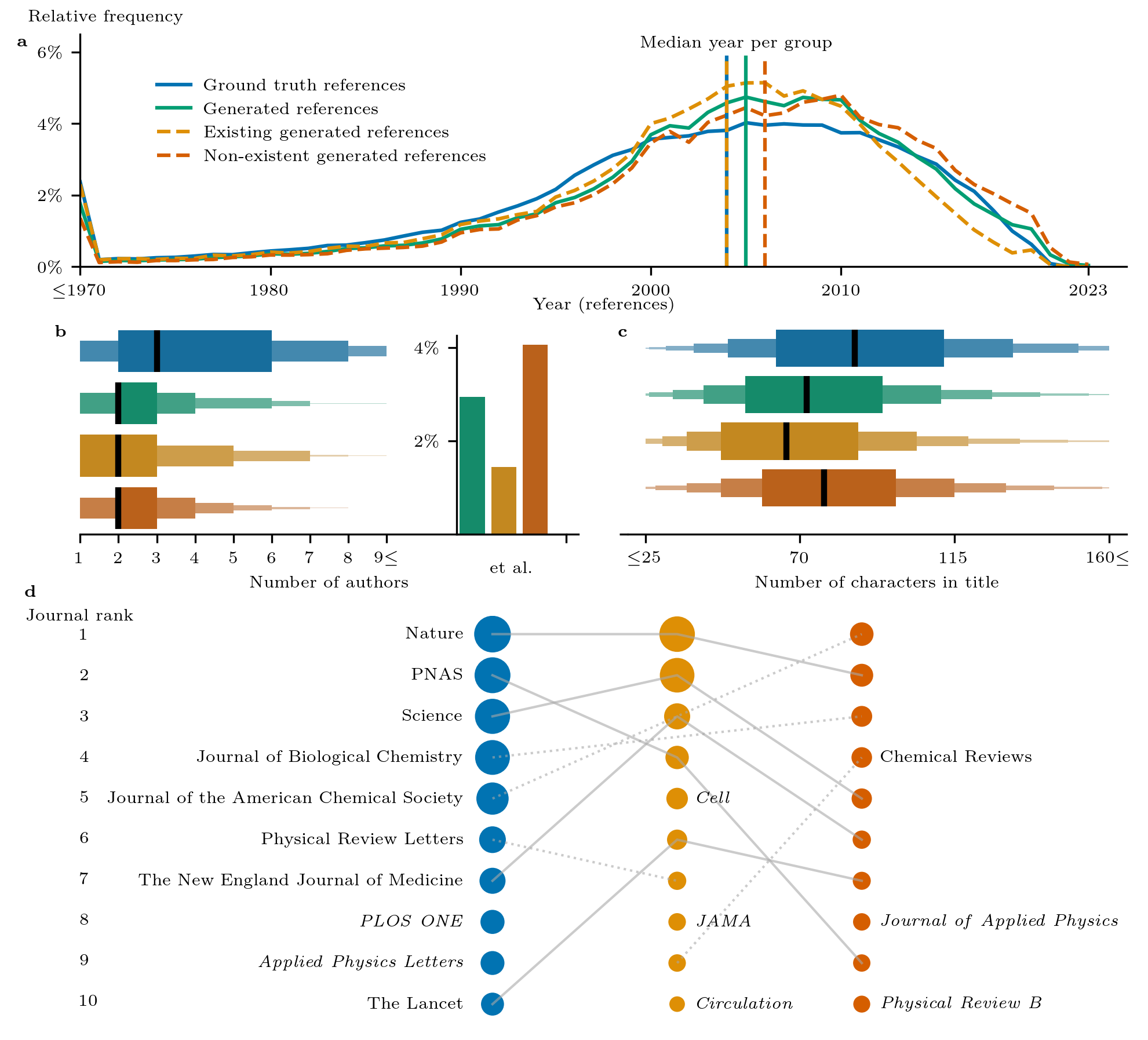}
	\caption{\textbf{Generated references exhibit a systematic preference for more recent references with shorter titles and fewer authors.} This figure summarizes key characteristics for ground truth (blue, n=$274,951$), generated (green, n=$274,497$), existing generated (orange, n=$116,939$), and non-existing generated (red, n=$157,558$) references. \textbf{a}, The relative frequency of publication years within each reference group, with median publication years indicated by vertical lines, shows that generated references are generally more recent than the ground truth. This recency bias is driven by non-existent generated references, which disproportionately cite more recent publications. Existing generated references show a more complex pattern, tempering the overall recency bias in the generated references. The pairwise two-sided Wilcoxon signed-rank test at the focal paper level confirms the statistically significant difference in median publication year between ground truth and generated references ($p$$<$$0.001$). \textbf{b}, The distribution of the number of authors shows that generated references tend to favor documents with fewer authors with a peak aroud 2-3 (1-3 for existing generated references) authors, compared to 2-6 authors for ground truth references.  A small proportion of generated references are labeled as ``et al.'' ($3\%$), with higher rates in non-existent ($4\%$) than existing generated references ($1.5\%$). The pairwise two-sided Wilcoxon signed-rank test at the focal paper level confirms the statistically significant difference in the median number of authors between ground truth and generated references ($p$$<$$0.001$). \textbf{c}, The distribution of the title length shows that generated references tend to favor documents with shorter titles. This effect is most outspoken for the existing generated references. The pairwise two-sided Wilcoxon signed-rank test at the focal paper level confirms the statistically significant difference in the median title length between ground truth and generated references ($p$$<$$0.001$). \textbf{d}, The journal rankings show the top 10 journals across different reference groups. The size of each dot represents how relatively frequently that journal appears within its reference group. Journals are connected by solid lines when appearing in all three groups' top 10, dotted lines when appearing in two groups' top 10, and shown in italic font when appearing in only one group's top 10, highlighting the distinct citation patterns across reference types.}
	\label{fig:main_4}
\end{figure}

\section*{Reinforcing the Matthew effect in citations}
Figure~\ref{fig:main_3} displays the existence rate, i.e., the fraction of generated references that corresponds to an existing reference in the SciSciNet \citep{lin2023sciscinet} database, of generated references (gray, n=$274,497$), and the citation characteristics of the ground truth (blue, n=$274,951$) and existing generated (orange, n=$116,939$) references across fields and time. Humanities and social sciences yield significantly higher existence rates than exact sciences (Figure~\ref{fig:main_3}\textbf{a}), though the overall rate remains stable between $40\%$ and $50\%$ across publication years of the focal papers (Figure~\ref{fig:main_3}\textbf{e}). The field-level variation persists even when controlling for the varying sample sizes of focal papers across fields through subsampling (Appendix Figure~\ref{fig:appendix_3}), and the existence rates on the focal paper level show only moderate negative correlation ($-0.10$, $p$$<$$0.001$) with the number of requested references (Appendix Figure~\ref{fig:appendix_2}). The higher existence rates in humanities and social sciences can be partially attributed to their tendency to cite older references both in absolute terms (Appendix Figure~\ref{fig:appendix_4}\textbf{b}) and relative to the publication year of the focal paper (Appendix Figure~\ref{fig:appendix_4}\textbf{c,d}), as existence rates drop significantly for more recently published references (Appendix Figure~\ref{fig:appendix_4}\textbf{a}). 

Existing generated references exhibit significantly higher median citation counts than ground truth references across all scientific fields (Figure~\ref{fig:main_3}\textbf{b}) and throughout all publication years of the focal papers (Figure~\ref{fig:main_3}\textbf{d}). This reinforcement of the Matthew effect in citations is confirmed through the pairwise two-sided Wilcoxon signed-rank tests at the focal paper level (see \hyperref[sec:appendix]{Appendix}). The observed gap in median citation counts between existing generated references and ground truth references remains even when selecting, for each focal paper, the same number of most highly cited ground truth references as the number of existing generated references. This confirms that the higher citation counts in existing generated references are not merely due to selection bias (Appendix Figure~\ref{fig:appendix_11}). The robustness of this effect is further strengthened by alternative citation metrics, including normalized citations \citep{radicchi2008universality} and citations after 5 and 10 years~\citep{wang2013citation} (Appendix Figure~\ref{fig:appendix_6}\textbf{a}), and the persistent gap in median citation counts across publication years of references (Appendix Figure~\ref{fig:appendix_6}\textbf{b}). The strong preference toward highly cited papers is also evidenced with approximately $90\%$ of existing generated references appearing in the top $10\%$ most cited papers within their respective field and year \citep{lin2023sciscinet}, and remarkably, over $60\%$ falling within the top $1\%$ -- more than twice the rate observed in ground truth references (Appendix Figure~\ref{fig:appendix_6}\textbf{c}).

Several factors may contribute to this citation disparity (Appendix Figure~\ref{fig:appendix_5}). The existing generated references have a higher proportion of books compared to ground truth references, which typically accumulate more citations. However, even the journal articles among the existing generated references achieve citation counts comparable to books in the ground truth references (Appendix Figure~\ref{fig:appendix_5}\textbf{a-c}). Moreover, the existing generated references demonstrate broader impact beyond traditional academic citations, exhibiting significantly higher rates of alternative citations such as patents and clinical trials, greater media attention through news coverage and social media mentions, and more frequent association with major funding sources like NIH and NSF grants (Appendix Figure~\ref{fig:appendix_5}\textbf{d}). Interestingly, despite these pronounced differences in citation impact, the reference counts between existing generated and ground truth papers show more moderate differences across fields, suggesting that the number of references is an unlikely source for the citation discrepancy \citep{mammola2021impact} (Figure~\ref{fig:main_3}\textbf{c}).

\section*{Systematic preferences in generated references}
Figure~\ref{fig:main_4} summarizes key characteristics, including publication year, number of authors, title length, and publication venue, for ground truth (blue, n=$274,951$), generated (green, n=$274,497$), existing generated (orange, n=$116,939$), and non-existing generated (red, n=$157,558$) references, indicating the model's internalization of citation patterns. The temporal distribution shows that generated references tend to be relatively more recent than ground truth references (Figure~\ref{fig:main_4}\textbf{a}), with the median difference being statistically significant as confirmed by a pairwise two-sided Wilcoxon signed-rank test at the focal paper level ($p$$<$$0.001$). The non-existent generated references amplify the recency bias documented in human citation behavior \citep{tahamtan2016factors}, while existing generated references show a more complex pattern: they under-cite earlier publications (pre-2000) compared to the ground truth, cite relatively more recent works between 2000 and 2012, and again under-cite in the most recent period, resulting in a median publication year equal to the ground truth. This temporal pattern of existing generated references suggests that their higher citation impact is not driven by a recency bias  (Figure~\ref{fig:main_3}\textbf{d}, Appendix Figure~\ref{fig:appendix_6}\textbf{b}). The concentration of references in the 2000s reflects both ground truth and generated references peaking for papers published within 0-5 years before the focal paper (Appendix Figure~\ref{fig:appendix_4}\textbf{c}), as publications from this period would be recent relative to many focal papers in our 1999-2021 sample.

The preference for smaller author teams emerges as another systematic pattern in generated references (Figure~\ref{fig:main_4}\textbf{b}). While ground truth references span author teams of 2-6 members, generated references concentrate on papers with 2-3 authors, and existing generated references focus even more narrowly on teams of 1-3 authors. This pattern, confirmed as statistically significant by a pairwise two-sided Wilcoxon signed-rank test at the focal paper level ($p$$<$$0.001$), stands in contrast to documented citation preferences for papers with larger author teams \citep{gazni2011investigating,thelwall2023coauthored}. The relatively small proportion of generated references labeled as ``et al.'' ($3\%$) is primarily driven by non-existent references ($4\%$), while existing generated references show a lower rate ($1.5\%$), suggesting that GPT-4o occasionally defaults to simplified author attribution when generating references that prove to be non-existent \citep{agrawal-etal-2024-language} (Figure\ref{fig:main_4}\textbf{b}).

Generated references show a clear preference for shorter titles (Figure~\ref{fig:main_4}\textbf{c}), with this effect being most pronounced in existing generated references. This pattern aligns with documented human citation behavior, where papers with shorter titles tend to receive more citations \citep{letchford2015advantage}. The statistically significant difference in median title length between ground truth and generated references ($p$$<$$0.001$) suggests that the model amplifies this human preference for concision. Together with the preference for smaller teams and the simplified author attribution, this tendency toward shorter titles may indicate a broader pattern of LLMs favoring simpler, more condensed bibliographic information. The two previous findings are exactly as expected, given that longer phrases are both harder to learn and to reproduce without error.

The analysis of publication venues reveals a strong preference for high-impact journals (Figure~\ref{fig:main_4}\textbf{d}). Consistent with documented human citation preferences for prestigious venues \citep{lawrence2003politics}, our findings show that generated references reinforce this citation pattern. The journal rankings demonstrate substantial overlap among leading multidisciplinary journals like Nature and Science, but also a notable concentration in top-tier field-specific journals such as The New England Journal of Medicine and Journal of the American Chemical Society. While PLOS ONE and Applied Physics Letters appear exclusively in the ground truth top 10, Cell, JAMA, and Circulation emerge uniquely in the existing generated references, suggesting that LLMs systematically favor references from the most prestigious venues within each field. Note that for the ground truth and existing generated references we can use the publication venue from the SciSciNet \citep{lin2023sciscinet} database, whereas for the non-existent generated references we use a validated text extraction method (Appendix Figure~\ref{fig:appendix_1}). Given the low representation of conference proceedings in our dataset (Appendix Figure~\ref{fig:appendix_5}\textbf{a}), we focus our analysis only on journal publications.

\section*{Network properties of generated references}
\begin{figure}[t!]
	\centering
	\includegraphics[width=1\textwidth]{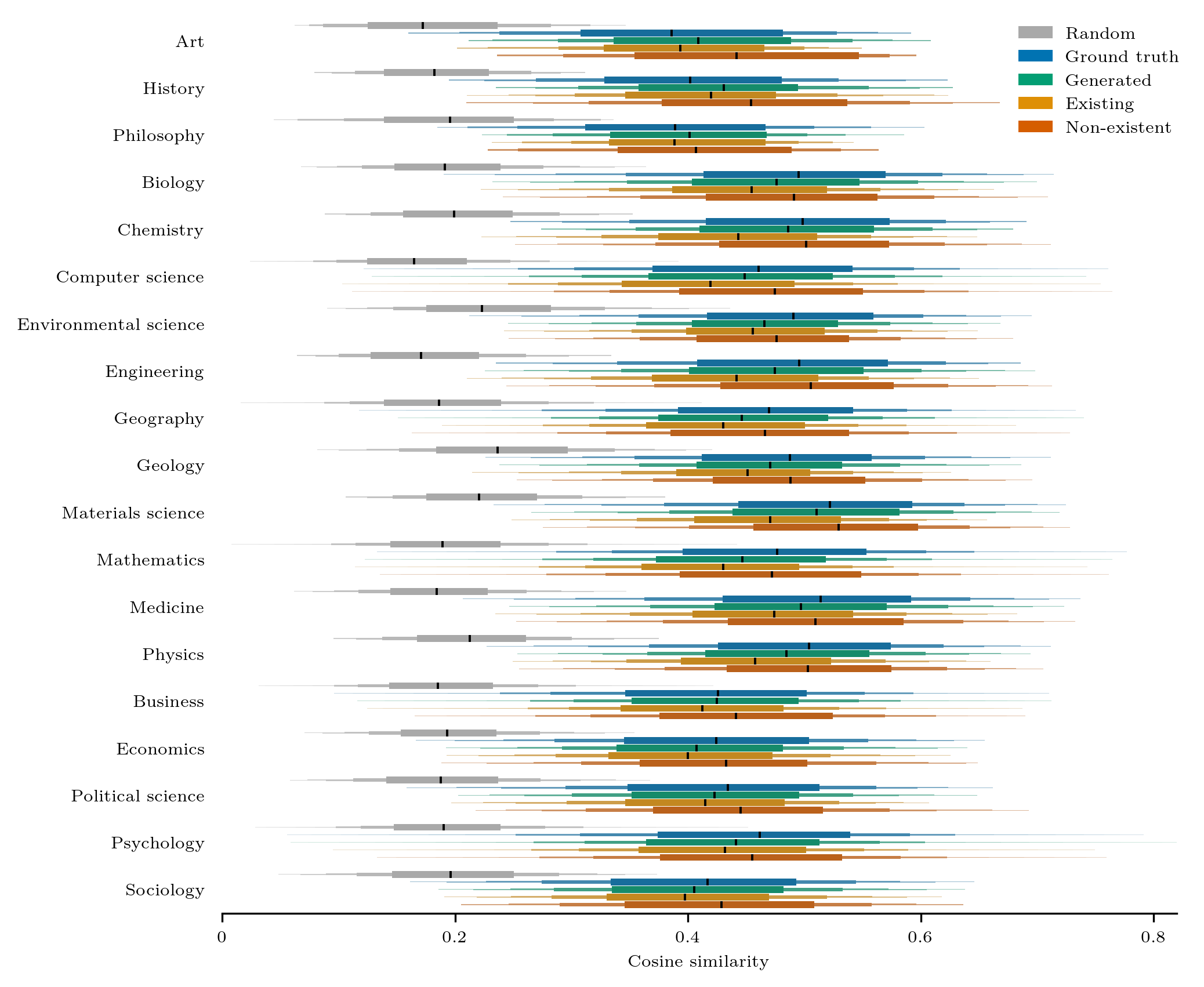}
\caption{\textbf{Generated references exhibit a level of cosine similarity to focal paper titles and abstracts on par with ground truth references, surpassing that of a random ground truth reference list from the same field.} This figure displays the distributions of the pairwise cosine similarity between OpenAI text-embedding-3-large vector embeddings (size=$3,072$) of the titles of the ground truth (blue, n=$274,951$), generated (green, n=$274,497$), existing generated (orange, n=$116,939$), and non-existing generated (red, n=$157,558$) references with the title and abstract of their corresponding focal paper (n=$10,000$). As a benchmark, we also compute for each focal paper the pairwise cosine similarity with the reference from a random ground truth reference list from the same field (gray, n=$274,951$).}
	\label{fig:main_5}
\end{figure}

Figure \ref{fig:main_5} shows the distributions of the pairwise cosine similarity between textual embeddings \citep{arts2023beyond,kusupati2022matryoshka,mobini2025deeply} of the titles of ground truth (blue, n=$274,951$), generated (green, n=$274,497$), existing generated (orange, n=$116,939$), and non-existing generated (red, n=$157,558$) references with the title and abstract of their corresponding focal paper (n=$10,000$). As a benchmark, we also compute for each focal paper the pairwise cosine similarity with the reference from a random ground truth reference list from the same field (gray, n=$274,951$). Notably, the non-existing generated references show slightly higher overall similarity scores than their existing counterparts, suggesting a tendency to overfit to the abstract and title by producing thematically aligned but non-existing references. However, all generated references, both existing and non-existent, remain substantially more content matched than random ground truth reference lists from the same field, underscoring the system's strong capacity for content alignment. We obtain qualitatively similar results with a different embeddings model (SPECTER2 \citep{singh-etal-2023-scirepeval}) and by only using the title of the focal paper (Appendix Figure~\ref{fig:appendix_8}). Interestingly, existing generated references exhibit significantly lower self-citation rates at the single-author level \citep{Szomszor2020self}, while still capturing the field-level variations observed in ground truth references (Appendix Figure~\ref{fig:appendix_7}).

Existing generated references exhibit notable overlap with human citation patterns, yet they do not perfectly replicate ground truth references at the corresponding focal paper level, suggesting that LLMs do not rely solely on direct memorization of citation lists~\citep{mobini2025deeply}. Only $8.78\%$ of existing generated references appear in the corresponding ground truth reference list, indicating that LLMs' reference generation is not driven by simple recall but rather by a broader pattern-matching process. However, when expanding the scope beyond corresponding focal papers, $22.24\%$ of existing generated references are found in at least one ground truth reference list across the dataset, suggesting that a substantial portion of LLM-generated references are genuinely cited in scientific literature. Additionally, $28.70\%$ of existing generated references overlap between different focal papers, revealing a preference for a recurring subset of frequently suggested references. This systematic overlap, combined with the strong preference for highly cited works, reinforces the notion that LLM-generated references amplify dominant citation patterns rather than diversifying the scientific discourse.

\begin{figure}[t!]
	\centering
	\includegraphics[width=1\textwidth]{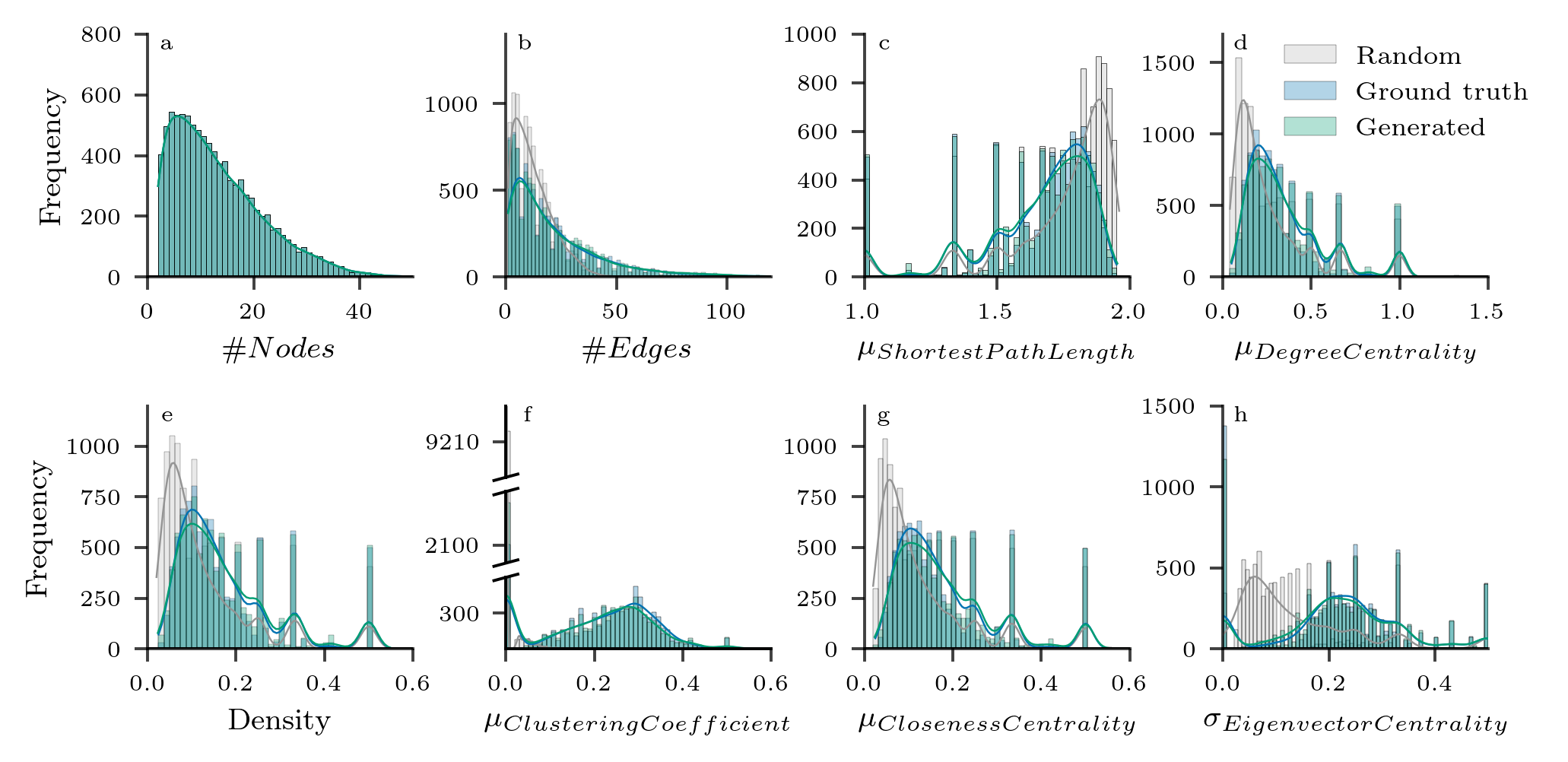}
\caption{\textbf{The citation graphs of the generated references are structurally similar to the citation graphs of the ground truth references, deviating significantly from random baselines.} This figure displays that the local citation graphs corresponding to the references generated by GPT4-o exhibit structural similarities with the local citation graphs corresponding to the ground truth references across several graph graph based measures, see the appendix for a detailed definitions of all the involved notions. Notably, these similarities cannot be found in local citation graphs corresponding to randomly reshuffled ground truth reference lists. For each of the $10,000$ focal papers, three local citation graphs were constructed. The graphs were based on the ground truth references of the focal paper, the GPT4-o generated references that exist (adding necessary edges to maintain connectivity for references not originally cited by the focal paper), and  a random baseline created by reshuffling the ground truth references (while preserving the field of study of the focal paper). Since approximately $50\%$ of the GPT4-o-generated references exist (see Fig. \ref{fig:main_3}\textbf{a}), the initial GPT4-o-based graphs had fewer nodes. To ensure a fair comparison, a random subset of nodes was removed from both the ground truth and random graphs, yielding three graphs of equal size per focal paper. All of these graphs were then converted to undirected graphs for analytical simplicity. A detailed description of the pipeline can be found in Appendix Figure \ref{fig:appendix_9}. \textbf{a,} The resulting node counts are identical for the GPT4-o (blue), ground truth (red), and random (gray) graphs at the focal paper level. \textbf{b,} The number of edges in GPT4-o-generated graphs closely aligns with human citation patterns, deviating significantly from the random baseline. \textbf{c,} GPT4-o-based citation graphs also mirror the distribution of average shortest path length in human citation networks, whereas the random graphs deviate substantially. \textbf{d–h}, The analogous observation for the mean degree centrality, density, average clustering coefficient, average closeness centrality, and standard deviation of the eigenvector centrality, highlighting GPT4-o's strong internalization of human citation patterns and second-order graph structures.}
	\label{fig:main_6}
\end{figure}

In Figure \ref{fig:main_6}, we investigate three distinct types of local citation graphs for each of the $10,000$ focal papers. Those graphs are based on the ground truth references, the (existing) GPT4-o-generated references, and randomly reshuffled ground truth references where we fix the field of study of the focal paper. Interestingly, we find that the graphs corresponding to the LLM-generated references strongly resemble human citation networks in terms of multiple graph-based measures, and deviate significantly from the random baseline, see Figure \ref{fig:main_6}\textbf{c-h}. This highlights the capacity of GPT4-o to internalize human citation patterns even on second-order graph structures~\citep{mobini2025deeply}. In Appendix Figure \ref{fig:appendix_9}, we also find that the existing generated references are much better connected to the ground truth references compared to the random baselines, both individually and on aggregate.

\section*{Discussion}
In summary, our findings reveal that LLMs do not merely replicate human citation practices but instead introduce systematic biases that reinforce dominant citation patterns. By analyzing $274,951$ references generated by GPT-4o for $10,000$ focal papers, we demonstrate that LLM-generated references exhibit a strong preference for highly cited papers, reinforcing the Matthew effect in citations. While the generated references align well with ground truth references in terms of content relevance, they systematically favor more recent works with shorter titles and fewer authors. Moreover, we find that LLM-generated references reduce author self-citations, suggesting that they deprioritize self-referential tendencies that are common in human citation behavior. Note that these biases and observed field-specific differences may partially stem from uneven representation of scientific works within the LLM's training corpus, especially affecting references from the long tail. Despite these differences, LLM-generated references maintain strong semantic alignment with focal papers, exhibiting cosine similarity scores comparable to human-curated citations across scientific fields. This tendency is even more pronounced for the local citation graphs corresponding to the (existing) generated references, resembling human citation networks across various graph-based measures compared to the random baseline. For both the embeddings and the graph based measures, the strong deviation from the random baseline, which essentially simulates randomly selecting references from the field of study of the focal paper, provides first evidence that LLMs do not merely suggest highly cited papers, but are capable of finding suitable references across various scientific domains solely by using their parametric knowledge.

Overall, our study highlights both the potential and the limitations of LLMs in shaping scientific citation practices. While these models capture broad citation trends, they may not fully reflect the nuanced citation behavior observed in human references. One key limitation of our approach is that it evaluates LLM-generated references in a controlled setting without interactive prompting or access to external databases, which would more accurately reflect real-world usage. Furthermore, by relying solely on parametric knowledge, our analysis isolates inherent biases in the model but does not account for the potential effects of retrieval-augmented generation. These constraints suggest that future research should explore how different LLM interactions, retrieval mechanisms, and training data compositions influence citation generation.

We open-source our dataset, extending the publicly available SciSciNet \citep{lin2023sciscinet}, thereby encouraging researchers interested in the intersection of LLM capabilities and the science of science to explore the potential impact of LLMs on scientific citation practices further. The implications of our findings extend beyond citation generation to broader discussions on automated scientific synthesis and knowledge dissemination. As LLMs become increasingly integrated into academic search engines, literature reviews, and automated research assistants, their influence on citation patterns could reshape how researchers discover and build upon prior work. Understanding these mechanisms is crucial for ensuring that AI-driven tools enhance, rather than distort, the trajectory of scientific discovery. Future work should assess how these models interact with real-world citation networks, peer review processes, and interdisciplinary knowledge flows, helping to guide the responsible deployment of LLMs in scientific research.

\clearpage
\section*{Acknowledgements}
Andres Algaba acknowledges a fellowship from the Research Foundation Flanders under Grant No.1286924N. Vincent Ginis acknowledges support from Research Foundation Flanders under Grant No.G032822N and G0K9322N. The resources and services used in this work were provided by the VSC (Flemish Supercomputer Center), funded by the Research Foundation - Flanders (FWO) and the Flemish Government.

\bibliographystyle{plain}
\bibliography{biblio}

\begin{thebibliography}{100}

\bibitem{acerbi2023large}
Alberto Acerbi and Joseph~M Stubbersfield.
\newblock Large language models show human-like content biases in transmission
  chain experiments.
\newblock {\em Proceedings of the National Academy of Sciences},
  120(44):e2313790120, 2023.

\bibitem{agrawal-etal-2024-language}
Ayush Agrawal, Mirac Suzgun, Lester Mackey, and Adam Kalai.
\newblock Do language models know when they{'}re hallucinating references?
\newblock In Yvette Graham and Matthew Purver, editors, {\em Findings of the
  Association for Computational Linguistics: EACL 2024}, pages 912--928, St.
  Julian{'}s, Malta, March 2024. Association for Computational Linguistics.

\bibitem{ajith2024litsearch}
Anirudh Ajith, Mengzhou Xia, Alexis Chevalier, Tanya Goyal, Danqi Chen, and
  Tianyu Gao.
\newblock Litsearch: A retrieval benchmark for scientific literature search.
\newblock {\em arXiv preprint arXiv:2407.18940}, 2024.

\bibitem{algaba2024largelanguagemodelsreflect}
Andres Algaba, Carmen Mazijn, Vincent Holst, Floriano Tori, Sylvia Wenmackers,
  and Vincent Ginis.
\newblock Large language models reflect human citation patterns with a
  heightened citation bias.
\newblock {\em arXiv preprint arXiv:2405.15739}, 2024.

\bibitem{arts2023beyond}
Sam Arts, Nicola Melluso, and Reinhilde Veugelers.
\newblock Beyond citations: Measuring novel scientific ideas and their impact
  in publication text.
\newblock {\em arXiv e-prints}, pages arXiv--2309, 2023.

\bibitem{baek2024researchagent}
Jinheon Baek, Sujay~Kumar Jauhar, Silviu Cucerzan, and Sung~Ju Hwang.
\newblock Researchagent: Iterative research idea generation over scientific
  literature with large language models.
\newblock {\em arXiv preprint arXiv:2404.07738}, 2024.

\bibitem{Bago2024science}
Bence Bago and Jean-François Bonnefon.
\newblock Generative ai as a tool for truth.
\newblock {\em Science}, 385(6714):1164--1165, 2024.

\bibitem{bick2024rapid}
Alexander Bick, Adam Blandin, and David~J Deming.
\newblock The rapid adoption of generative ai.
\newblock {\em National Bureau of Economic Research}, 2024.

\bibitem{boiko2023emergent}
Daniil~A Boiko, Robert MacKnight, and Gabe Gomes.
\newblock Emergent autonomous scientific research capabilities of large
  language models.
\newblock {\em arXiv preprint arXiv:2304.05332}, 2023.

\bibitem{bornmann2008citation}
Lutz Bornmann and Hans-Dieter Daniel.
\newblock What do citation counts measure? a review of studies on citing
  behavior.
\newblock {\em Journal of documentation}, 64(1):45--80, 2008.

\bibitem{brown2020language}
Tom Brown, Benjamin Mann, Nick Ryder, Melanie Subbiah, Jared~D Kaplan, Prafulla
  Dhariwal, Arvind Neelakantan, Pranav Shyam, Girish Sastry, Amanda Askell,
  et~al.
\newblock Language models are few-shot learners.
\newblock {\em Advances in neural information processing systems},
  33:1877--1901, 2020.

\bibitem{bubeck2023sparks}
S{\'e}bastien Bubeck, Varun Chandrasekaran, Ronen Eldan, Johannes Gehrke, Eric
  Horvitz, Ece Kamar, Peter Lee, Yin~Tat Lee, Yuanzhi Li, Scott Lundberg,
  et~al.
\newblock Sparks of artificial general intelligence: Early experiments with
  gpt-4.
\newblock {\em arXiv preprint arXiv:2303.12712}, 2023.

\bibitem{checco2021ai}
Alessandro Checco, Lorenzo Bracciale, Pierpaolo Loreti, Stephen Pinfield, and
  Giuseppe Bianchi.
\newblock Ai-assisted peer review.
\newblock {\em Humanities and social sciences communications}, 8(1):1--11,
  2021.

\bibitem{Chen_Lin_Han_Sun_2024}
Jiawei Chen, Hongyu Lin, Xianpei Han, and Le~Sun.
\newblock Benchmarking large language models in retrieval-augmented generation.
\newblock {\em Proceedings of the AAAI Conference on Artificial Intelligence},
  38(16):17754--17762, Mar. 2024.

\bibitem{chen2021evaluating}
Mark Chen, Jerry Tworek, Heewoo Jun, Qiming Yuan, Henrique Ponde De~Oliveira
  Pinto, Jared Kaplan, Harri Edwards, Yuri Burda, Nicholas Joseph, Greg
  Brockman, et~al.
\newblock Evaluating large language models trained on code.
\newblock {\em arXiv preprint arXiv:2107.03374}, 2021.

\bibitem{chen2025unveiling}
Xiuying Chen, Tairan Wang, Taicheng Guo, Kehan Guo, Juexiao Zhou, Haoyang Li,
  Zirui Song, Xin Gao, and Xiangliang Zhang.
\newblock Unveiling the power of language models in chemical research question
  answering.
\newblock {\em Communications Chemistry}, 8(1):4, 2025.

\bibitem{christen2012data}
Peter Christen and Peter Christen.
\newblock {\em The data matching process}.
\newblock Springer, 2012.

\bibitem{clark2018think}
Peter Clark, Isaac Cowhey, Oren Etzioni, Tushar Khot, Ashish Sabharwal, Carissa
  Schoenick, and Oyvind Tafjord.
\newblock Think you have solved question answering? try arc, the ai2 reasoning
  challenge.
\newblock {\em arXiv preprint arXiv:1803.05457}, 2018.

\bibitem{cobbe2021training}
Karl Cobbe, Vineet Kosaraju, Mohammad Bavarian, Mark Chen, Heewoo Jun, Lukasz
  Kaiser, Matthias Plappert, Jerry Tworek, Jacob Hilton, Reiichiro Nakano,
  et~al.
\newblock Training verifiers to solve math word problems.
\newblock {\em arXiv preprint arXiv:2110.14168}, 2021.

\bibitem{crossref_api}
CrossRef.
\newblock \url{https://www.crossref.org/}, 2024.

\bibitem{delgado2025review}
Fernando~M. Delgado-Chaves, Matthew~J. Jennings, Antonio Atalaia, Justus Wolff,
  Rita Horvath, Zeinab~M. Mamdouh, Jan Baumbach, and Linda Baumbach.
\newblock Transforming literature screening: The emerging role of large
  language models in systematic reviews.
\newblock {\em Proceedings of the National Academy of Sciences},
  122(2):e2411962122, 2025.

\bibitem{dennstadt2024title}
Fabio Dennst{\"a}dt, Johannes Zink, Paul~Martin Putora, Janna Hastings, and
  Nikola Cihoric.
\newblock Title and abstract screening for literature reviews using large
  language models: an exploratory study in the biomedical domain.
\newblock {\em Systematic Reviews}, 13(1):158, 2024.

\bibitem{dubey2024llama}
Abhimanyu Dubey, Abhinav Jauhri, Abhinav Pandey, Abhishek Kadian, Ahmad
  Al-Dahle, Aiesha Letman, Akhil Mathur, Alan Schelten, Amy Yang, Angela Fan,
  et~al.
\newblock The llama 3 herd of models.
\newblock {\em arXiv preprint arXiv:2407.21783}, 2024.

\bibitem{elasticsearch}
Elastic.
\newblock Elasticsearch.
\newblock \url{https://github.com/elastic/elasticsearch}, 2024.
\newblock Free and Open Source, Distributed, RESTful Search Engine.

\bibitem{evans2008electronic}
James~A Evans.
\newblock Electronic publication and the narrowing of science and scholarship.
\newblock {\em science}, 321(5887):395--399, 2008.

\bibitem{fabiano2024optimize}
Nicholas Fabiano, Arnav Gupta, Nishaant Bhambra, Brandon Luu, Stanley Wong,
  Muhammad Maaz, Jess~G Fiedorowicz, Andrew~L Smith, and Marco Solmi.
\newblock How to optimize the systematic review process using ai tools.
\newblock {\em JCPP Advances}, page e12234, 2024.

\bibitem{fortunato2018science}
Santo Fortunato, Carl~T Bergstrom, Katy B{\"o}rner, James~A Evans, Dirk
  Helbing, Sta{\v{s}}a Milojevi{\'c}, Alexander~M Petersen, Filippo Radicchi,
  Roberta Sinatra, Brian Uzzi, et~al.
\newblock Science of science.
\newblock {\em Science}, 359(6379):eaao0185, 2018.

\bibitem{gao2023retrieval}
Tianyu Gao, Howard Yen, Jiatong Yu, and Danqi Chen.
\newblock Enabling large language models to generate text with citations.
\newblock {\em arXiv preprint arXiv:2305.14627}, 2023.

\bibitem{gazni2011investigating}
Ali Gazni and Fereshteh Didegah.
\newblock Investigating different types of research collaboration and citation
  impact: a case study of harvard university's publications.
\newblock {\em Scientometrics}, 87(2):251--265, 2011.

\bibitem{ghafarollahi2024sciagents}
Alireza Ghafarollahi and Markus~J Buehler.
\newblock Sciagents: Automating scientific discovery through multi-agent
  intelligent graph reasoning.
\newblock {\em arXiv preprint arXiv:2409.05556}, 2024.

\bibitem{gottweis2025towards}
Juraj Gottweis, Wei-Hung Weng, Alexander Daryin, Tao Tu, Anil Palepu, Petar
  Sirkovic, Artiom Myaskovsky, Felix Weissenberger, Keran Rong, Ryutaro Tanno,
  et~al.
\newblock Towards an ai co-scientist.
\newblock {\em arXiv preprint arXiv:2502.18864}, 2025.

\bibitem{gougherty2024testing}
Andrew~V Gougherty and Hannah~L Clipp.
\newblock Testing the reliability of an ai-based large language model to
  extract ecological information from the scientific literature.
\newblock {\em npj Biodiversity}, 3(1):13, 2024.

\bibitem{greenberg2009citation}
Steven~A Greenberg.
\newblock How citation distortions create unfounded authority: analysis of a
  citation network.
\newblock {\em Bmj}, 339, 2009.

\bibitem{guo2025deepseek}
Daya Guo, Dejian Yang, Haowei Zhang, Junxiao Song, Ruoyu Zhang, Runxin Xu,
  Qihao Zhu, Shirong Ma, Peiyi Wang, Xiao Bi, et~al.
\newblock Deepseek-r1: Incentivizing reasoning capability in llms via
  reinforcement learning.
\newblock {\em arXiv preprint arXiv:2501.12948}, 2025.

\bibitem{gurnee2024languagemodelsrepresentspace}
Wes Gurnee and Max Tegmark.
\newblock Language models represent space and time.
\newblock {\em arXiv preprint arXiv:2310.02207}, 2024.

\bibitem{han2024mining}
Sukjin Han.
\newblock Mining causality: Ai-assisted search for instrumental variables.
\newblock {\em arXiv preprint arXiv:2409.14202}, 2024.

\bibitem{hao2023reasoning}
Shibo Hao, Yi~Gu, Haodi Ma, Joshua~Jiahua Hong, Zhen Wang, Daisy~Zhe Wang, and
  Zhiting Hu.
\newblock Reasoning with language model is planning with world model.
\newblock {\em arXiv preprint arXiv:2305.14992}, 2023.

\bibitem{hendrycks2020measuring}
Dan Hendrycks, Collin Burns, Steven Basart, Andy Zou, Mantas Mazeika, Dawn
  Song, and Jacob Steinhardt.
\newblock Measuring massive multitask language understanding.
\newblock {\em arXiv preprint arXiv:2009.03300}, 2020.

\bibitem{hendrycks2021measuring}
Dan Hendrycks, Collin Burns, Saurav Kadavath, Akul Arora, Steven Basart, Eric
  Tang, Dawn Song, and Jacob Steinhardt.
\newblock Measuring mathematical problem solving with the math dataset.
\newblock {\em arXiv preprint arXiv:2103.03874}, 2021.

\bibitem{hewitt2024predictions}
Luke Hewitt, Ashwini Ashokkumar, Isaias Ghezae, and Robb Willer.
\newblock Predicting results of social science experiments using large language
  models.
\newblock 2024.

\bibitem{hollmann2023automated}
Noah Hollmann, Samuel M\"{u}ller, and Frank Hutter.
\newblock Large language models for automated data science: Introducing caafe
  for context-aware automated feature engineering.
\newblock In A.~Oh, T.~Naumann, A.~Globerson, K.~Saenko, M.~Hardt, and
  S.~Levine, editors, {\em Advances in Neural Information Processing Systems},
  volume~36, pages 44753--44775. Curran Associates, Inc., 2023.

\bibitem{holst2024dataset}
Vincent Holst, Andres Algaba, Floriano Tori, Sylvia Wenmackers, and Vincent
  Ginis.
\newblock Dataset artefacts are the hidden drivers of the declining
  disruptiveness in science.
\newblock {\em arXiv preprint arXiv:2402.14583}, 2024.

\bibitem{horbach2022automated}
Serge~PJM Horbach, Freek~JW Oude~Maatman, Willem Halffman, and Wytske~M
  Hepkema.
\newblock Automated citation recommendation tools encourage questionable
  citations.
\newblock {\em Research Evaluation}, 31(3):321--325, 2022.

\bibitem{huang2023survey}
Lei Huang, Weijiang Yu, Weitao Ma, Weihong Zhong, Zhangyin Feng, Haotian Wang,
  Qianglong Chen, Weihua Peng, Xiaocheng Feng, Bing Qin, et~al.
\newblock A survey on hallucination in large language models: Principles,
  taxonomy, challenges, and open questions.
\newblock {\em arXiv preprint arXiv:2311.05232}, 2023.

\bibitem{jimenez2024swebench}
Carlos~E Jimenez, John Yang, Alexander Wettig, Shunyu Yao, Kexin Pei, Ofir
  Press, and Karthik~R Narasimhan.
\newblock {SWE}-bench: Can language models resolve real-world github issues?
\newblock In {\em The Twelfth International Conference on Learning
  Representations}, 2024.

\bibitem{kadavath2022language}
Saurav Kadavath, Tom Conerly, Amanda Askell, Tom Henighan, Dawn Drain, Ethan
  Perez, Nicholas Schiefer, Zac Hatfield-Dodds, Nova DasSarma, Eli
  Tran-Johnson, et~al.
\newblock Language models (mostly) know what they know.
\newblock {\em arXiv preprint arXiv:2207.05221}, 2022.

\bibitem{kandpal2023longtail}
Nikhil Kandpal, Haikang Deng, Adam Roberts, Eric Wallace, and Colin Raffel.
\newblock Large language models struggle to learn long-tail knowledge.
\newblock In {\em International Conference on Machine Learning}, pages
  15696--15707. PMLR, 2023.

\bibitem{kang2024researcharena}
Hao Kang and Chenyan Xiong.
\newblock Researcharena: Benchmarking llms' ability to collect and organize
  information as research agents.
\newblock {\em arXiv preprint arXiv:2406.10291}, 2024.

\bibitem{kinney2023semantic}
Rodney Kinney, Chloe Anastasiades, Russell Authur, Iz~Beltagy, Jonathan Bragg,
  Alexandra Buraczynski, Isabel Cachola, Stefan Candra, Yoganand Chandrasekhar,
  Arman Cohan, et~al.
\newblock The semantic scholar open data platform.
\newblock {\em arXiv preprint arXiv:2301.10140}, 2023.

\bibitem{kojima2022zeroshot}
Takeshi Kojima, Shixiang~Shane Gu, Machel Reid, Yutaka Matsuo, and Yusuke
  Iwasawa.
\newblock Large language models are zero-shot reasoners.
\newblock {\em Advances in neural information processing systems},
  35:22199--22213, 2022.

\bibitem{kusupati2022matryoshka}
Aditya Kusupati, Gantavya Bhatt, Aniket Rege, Matthew Wallingford, Aditya
  Sinha, Vivek Ramanujan, William Howard-Snyder, Kaifeng Chen, Sham Kakade,
  Prateek Jain, et~al.
\newblock Matryoshka representation learning.
\newblock {\em Advances in Neural Information Processing Systems},
  35:30233--30249, 2022.

\bibitem{mattheweffect}
Vincent Larivière and Yves Gingras.
\newblock The impact factor's matthew effect: A natural experiment in
  bibliometrics.
\newblock {\em Journal of the American Society for Information Science and
  Technology}, 61(2):424--427, 2010.

\bibitem{lawrence2003politics}
Peter~A Lawrence.
\newblock The politics of publication.
\newblock {\em Nature}, 422(6929):259--261, 2003.

\bibitem{lehr2024chatgpt}
Steven~A. Lehr, Aylin Caliskan, Suneragiri Liyanage, and Mahzarin~R. Banaji.
\newblock Chatgpt as research scientist: Probing gpt’s capabilities as a
  research librarian, research ethicist, data generator, and data predictor.
\newblock {\em Proceedings of the National Academy of Sciences},
  121(35):e2404328121, 2024.

\bibitem{letchford2015advantage}
Adrian Letchford, Helen~Susannah Moat, and Tobias Preis.
\newblock The advantage of short paper titles.
\newblock {\em Royal Society open science}, 2(8):150266, 2015.

\bibitem{lewis2020retrieval}
Patrick Lewis, Ethan Perez, Aleksandra Piktus, Fabio Petroni, Vladimir
  Karpukhin, Naman Goyal, Heinrich K{\"u}ttler, Mike Lewis, Wen-tau Yih, Tim
  Rockt{\"a}schel, et~al.
\newblock Retrieval-augmented generation for knowledge-intensive nlp tasks.
\newblock {\em Advances in Neural Information Processing Systems},
  33:9459--9474, 2020.

\bibitem{li2024scilitllm}
Sihang Li, Jin Huang, Jiaxi Zhuang, Yaorui Shi, Xiaochen Cai, Mingjun Xu, Xiang
  Wang, Linfeng Zhang, Guolin Ke, and Hengxing Cai.
\newblock Scilitllm: How to adapt llms for scientific literature understanding.
\newblock {\em arXiv preprint arXiv:2408.15545}, 2024.

\bibitem{lin2023sciscinet}
Zihang Lin, Yian Yin, Lu~Liu, and Dashun Wang.
\newblock Sciscinet: A large-scale open data lake for the science of science
  research.
\newblock {\em Scientific Data}, 10(1):315, 2023.

\bibitem{liu2023review}
Lu~Liu, Benjamin~F. Jones, Brian Uzzi, and Dashun Wang.
\newblock Data, measurement and empirical methods in the science of science.
\newblock {\em Nature Human Behaviour}, 7(7):1046--1058, 2023.

\bibitem{lu2024aiscientist}
Chris Lu, Cong Lu, Robert~Tjarko Lange, Jakob Foerster, Jeff Clune, and David
  Ha.
\newblock The ai scientist: Towards fully automated open-ended scientific
  discovery.
\newblock {\em arXiv preprint arXiv:2408.06292}, 2024.

\bibitem{m2024augmenting}
Andres M.~Bran, Sam Cox, Oliver Schilter, Carlo Baldassari, Andrew~D White, and
  Philippe Schwaller.
\newblock Augmenting large language models with chemistry tools.
\newblock {\em Nature Machine Intelligence}, pages 1--11, 2024.

\bibitem{fairprompting}
Huan Ma, Changqing Zhang, Yatao Bian, Lemao Liu, Zhirui Zhang, Peilin Zhao, Shu
  Zhang, Huazhu Fu, Qinghua Hu, and Bingzhe Wu.
\newblock Fairness-guided few-shot prompting for large language models.
\newblock In A.~Oh, T.~Naumann, A.~Globerson, K.~Saenko, M.~Hardt, and
  S.~Levine, editors, {\em Advances in Neural Information Processing Systems},
  volume~36, pages 43136--43155. Curran Associates, Inc., 2023.

\bibitem{machlup1980knowledge}
Fritz Machlup.
\newblock {\em Knowledge: Its Creation, Distribution, and Economic
  Significance, Volume I: Knowledge and Knowledge Production}.
\newblock Princeton University Press, Princeton, NJ, 1980.

\bibitem{mammola2021impact}
Stefano Mammola, Diego Fontaneto, Alejandro Mart{\'\i}nez, and Filipe
  Chichorro.
\newblock Impact of the reference list features on the number of citations.
\newblock {\em Scientometrics}, 126:785--799, 2021.

\bibitem{manning2024automatedsocialsciencelanguage}
Benjamin~S. Manning, Kehang Zhu, and John~J. Horton.
\newblock Automated social science: Language models as scientist and subjects.
\newblock {\em arXiv preprint arXiv:2404.11794}, 2024.

\bibitem{merchant2023scaling}
Amil Merchant, Simon Batzner, Samuel~S Schoenholz, Muratahan Aykol, Gowoon
  Cheon, and Ekin~Dogus Cubuk.
\newblock Scaling deep learning for materials discovery.
\newblock {\em Nature}, 624(7990):80--85, 2023.

\bibitem{mishra2024use}
Tanisha Mishra, Edward Sutanto, Rini Rossanti, Nayana Pant, Anum Ashraf, Akshay
  Raut, Germaine Uwabareze, Ajayi Oluwatomiwa, and Bushra Zeeshan.
\newblock Use of large language models as artificial intelligence tools in
  academic research and publishing among global clinical researchers.
\newblock {\em Scientific Reports}, 14(1):31672, 2024.

\bibitem{mobini2025deeply}
Melika Mobini, Vincent Holst, Floriano Tori, Andres Algaba, and Vincent Ginis.
\newblock How deeply do llms internalize human citation practices? a
  graph-structural and embedding-based evaluation.
\newblock In {\em ICLR 2025 Workshop on Human-AI Coevolution}, 2025.

\bibitem{muennighoff2025s1simpletesttimescaling}
Niklas Muennighoff, Zitong Yang, Weijia Shi, Xiang~Lisa Li, Li~Fei-Fei,
  Hannaneh Hajishirzi, Luke Zettlemoyer, Percy Liang, Emmanuel Candès, and
  Tatsunori Hashimoto.
\newblock s1: Simple test-time scaling.
\newblock {\em arXiv preprint arXiv:2501.19393}, 2025.

\bibitem{navigli2023biases}
Roberto Navigli, Simone Conia, and Bj{\"o}rn Ross.
\newblock Biases in large language models: origins, inventory, and discussion.
\newblock {\em ACM Journal of Data and Information Quality}, 15(2):1--21, 2023.

\bibitem{nielsen2021global}
Mathias~Wullum Nielsen and Jens~Peter Andersen.
\newblock Global citation inequality is on the rise.
\newblock {\em Proceedings of the National Academy of Sciences},
  118(7):e2012208118, 2021.

\bibitem{phan2025humanity}
Long Phan, Alice Gatti, Ziwen Han, Nathaniel Li, Josephina Hu, Hugh Zhang, Sean
  Shi, Michael Choi, Anish Agrawal, Arnav Chopra, et~al.
\newblock Humanity's last exam.
\newblock {\em arXiv preprint arXiv:2501.14249}, 2025.

\bibitem{press2024citeme}
Ori Press, Andreas Hochlehnert, Ameya Prabhu, Vishaal Udandarao, Ofir Press,
  and Matthias Bethge.
\newblock Cite{ME}: Can language models accurately cite scientific claims?
\newblock In {\em The Thirty-eight Conference on Neural Information Processing
  Systems Datasets and Benchmarks Track}, 2024.

\bibitem{price1965networks}
Derek J De~Solla Price.
\newblock Networks of scientific papers: The pattern of bibliographic
  references indicates the nature of the scientific research front.
\newblock {\em Science}, 149(3683):510--515, 1965.

\bibitem{qureshi2023chatgpt}
Riaz Qureshi, Daniel Shaughnessy, Kayden~AR Gill, Karen~A Robinson, Tianjing
  Li, and Eitan Agai.
\newblock Are chatgpt and large language models ``the answer'' to bringing us
  closer to systematic review automation?
\newblock {\em Systematic Reviews}, 12(1):72, 2023.

\bibitem{radicchi2008universality}
Filippo Radicchi, Santo Fortunato, and Claudio Castellano.
\newblock Universality of citation distributions: Toward an objective measure
  of scientific impact.
\newblock {\em Proceedings of the National Academy of Sciences},
  105(45):17268--17272, 2008.

\bibitem{rein2023gpqa}
David Rein, Betty~Li Hou, Asa~Cooper Stickland, Jackson Petty, Richard~Yuanzhe
  Pang, Julien Dirani, Julian Michael, and Samuel~R Bowman.
\newblock Gpqa: A graduate-level google-proof q\&a benchmark.
\newblock {\em arXiv preprint arXiv:2311.12022}, 2023.

\bibitem{romera2024mathematical}
Bernardino Romera-Paredes, Mohammadamin Barekatain, Alexander Novikov, Matej
  Balog, M~Pawan Kumar, Emilien Dupont, Francisco~JR Ruiz, Jordan~S Ellenberg,
  Pengming Wang, Omar Fawzi, et~al.
\newblock Mathematical discoveries from program search with large language
  models.
\newblock {\em Nature}, 625(7995):468--475, 2024.

\bibitem{schmidgall2025agentlaboratoryusingllm}
Samuel Schmidgall, Yusheng Su, Ze~Wang, Ximeng Sun, Jialian Wu, Xiaodong Yu,
  Jiang Liu, Zicheng Liu, and Emad Barsoum.
\newblock Agent laboratory: Using llm agents as research assistants.
\newblock {\em arXiv preprint arXiv:2501.04227}, 2025.

\bibitem{schneider2024something}
Jesper~W Schneider, Nick Allum, Jens~Peter Andersen, Michael~Bang Petersen,
  Emil~B Madsen, Niels Mejlgaard, and Robert Zachariae.
\newblock Is something rotten in the state of denmark? cross-national evidence
  for widespread involvement but not systematic use of questionable research
  practices across all fields of research.
\newblock {\em PloS one}, 19(8):e0304342, 2024.

\bibitem{schutze2008introduction}
Hinrich Sch{\"u}tze, Christopher~D Manning, and Prabhakar Raghavan.
\newblock {\em Introduction to information retrieval}, volume~39.
\newblock Cambridge University Press Cambridge, 2008.

\bibitem{si2024can}
Chenglei Si, Diyi Yang, and Tatsunori Hashimoto.
\newblock Can llms generate novel research ideas? a large-scale human study
  with 100+ nlp researchers.
\newblock {\em arXiv preprint arXiv:2409.04109}, 2024.

\bibitem{simkin2002read}
Mikhail~V Simkin and Vwani~P Roychowdhury.
\newblock Read before you cite!
\newblock {\em arXiv preprint cond-mat/0212043}, 2002.

\bibitem{singh-etal-2023-scirepeval}
Amanpreet Singh, Mike D{'}Arcy, Arman Cohan, Doug Downey, and Sergey Feldman.
\newblock {S}ci{R}ep{E}val: A multi-format benchmark for scientific document
  representations.
\newblock In Houda Bouamor, Juan Pino, and Kalika Bali, editors, {\em
  Proceedings of the 2023 Conference on Empirical Methods in Natural Language
  Processing}, pages 5548--5566, Singapore, December 2023. Association for
  Computational Linguistics.

\bibitem{skarlinski2024language}
Michael~D Skarlinski, Sam Cox, Jon~M Laurent, James~D Braza, Michaela Hinks,
  Michael~J Hammerling, Manvitha Ponnapati, Samuel~G Rodriques, and Andrew~D
  White.
\newblock Language agents achieve superhuman synthesis of scientific knowledge.
\newblock {\em arXiv preprint arXiv:2409.13740}, 2024.

\bibitem{skitka1999does}
Linda~J Skitka, Kathleen~L Mosier, and Mark Burdick.
\newblock Does automation bias decision-making?
\newblock {\em International Journal of Human-Computer Studies},
  51(5):991--1006, 1999.

\bibitem{smith2012impact}
Derek~R Smith.
\newblock Impact factors, scientometrics and the history of citation-based
  research.
\newblock {\em Scientometrics}, 92(2):419--427, 2012.

\bibitem{srivastava2022beyond}
Aarohi Srivastava, Abhinav Rastogi, Abhishek Rao, Abu Awal~Md Shoeb, Abubakar
  Abid, Adam Fisch, Adam~R Brown, Adam Santoro, Aditya Gupta, Adri{\`a}
  Garriga-Alonso, et~al.
\newblock Beyond the imitation game: Quantifying and extrapolating the
  capabilities of language models.
\newblock {\em arXiv preprint arXiv:2206.04615}, 2022.

\bibitem{susnjak2024automating}
Teo Susnjak, Peter Hwang, Napoleon~H Reyes, Andre~LC Barczak, Timothy~R
  McIntosh, and Surangika Ranathunga.
\newblock Automating research synthesis with domain-specific large language
  model fine-tuning.
\newblock {\em arXiv preprint arXiv:2404.08680}, 2024.

\bibitem{swanson2024virtual}
Kyle Swanson, Wesley Wu, Nash~L Bulaong, John~E Pak, and James Zou.
\newblock The virtual lab: Ai agents design new sars-cov-2 nanobodies with
  experimental validation.
\newblock {\em bioRxiv}, pages 2024--11, 2024.

\bibitem{Szomszor2020self}
Martin Szomszor, David~A. Pendlebury, and Jonathan Adams.
\newblock How much is too much? the difference between research influence and
  self-citation excess.
\newblock {\em Scientometrics}, 123(2):1119--1147, 2020.

\bibitem{tahamtan2018core}
Iman Tahamtan and Lutz Bornmann.
\newblock Core elements in the process of citing publications: Conceptual
  overview of the literature.
\newblock {\em Journal of informetrics}, 12(1):203--216, 2018.

\bibitem{tahamtan2016factors}
Iman Tahamtan, Askar Safipour~Afshar, and Khadijeh Ahamdzadeh.
\newblock Factors affecting number of citations: a comprehensive review of the
  literature.
\newblock {\em Scientometrics}, 107:1195--1225, 2016.

\bibitem{tam2024let}
Zhi~Rui Tam, Cheng-Kuang Wu, Yi-Lin Tsai, Chieh-Yen Lin, Hung-yi Lee, and
  Yun-Nung Chen.
\newblock Let me speak freely? a study on the impact of format restrictions on
  performance of large language models.
\newblock {\em arXiv preprint arXiv:2408.02442}, 2024.

\bibitem{team2023gemini}
Gemini Team, Rohan Anil, Sebastian Borgeaud, Yonghui Wu, Jean-Baptiste Alayrac,
  Jiahui Yu, Radu Soricut, Johan Schalkwyk, Andrew~M Dai, Anja Hauth, et~al.
\newblock Gemini: a family of highly capable multimodal models.
\newblock {\em arXiv preprint arXiv:2312.11805}, 2023.

\bibitem{thelwall2023coauthored}
Mike Thelwall, Kayvan Kousha, Mahshid Abdoli, Emma Stuart, Meiko Makita, Paul
  Wilson, and Jonathan Levitt.
\newblock Why are coauthored academic articles more cited: Higher quality or
  larger audience?
\newblock {\em Journal of the Association for Information Science and
  Technology}, 74(7):791--810, 2023.

\bibitem{tjuatja2024llms}
Lindia Tjuatja, Valerie Chen, Tongshuang Wu, Ameet Talwalkwar, and Graham
  Neubig.
\newblock Do llms exhibit human-like response biases? a case study in survey
  design.
\newblock {\em Transactions of the Association for Computational Linguistics},
  12:1011--1026, 2024.

\bibitem{trinh2024solving}
Trieu~H Trinh, Yuhuai Wu, Quoc~V Le, He~He, and Thang Luong.
\newblock Solving olympiad geometry without human demonstrations.
\newblock {\em Nature}, 625(7995):476--482, 2024.

\bibitem{Truhn2023reasoning}
Daniel Truhn, Jorge~S. Reis-Filho, and Jakob~Nikolas Kather.
\newblock Large language models should be used as scientific reasoning engines,
  not knowledge databases.
\newblock {\em Nature Medicine}, 29(12):2983--2984, 2023.

\bibitem{Trujillo2018cocitations}
Caleb~M. Trujillo and Tammy~M. Long.
\newblock Document co-citation analysis to enhance transdisciplinary research.
\newblock {\em Science Advances}, 4(1):e1701130, 2018.

\bibitem{walters2023fabrication}
William~H Walters and Esther~Isabelle Wilder.
\newblock Fabrication and errors in the bibliographic citations generated by
  chatgpt.
\newblock {\em Scientific Reports}, 13(1):14045, 2023.

\bibitem{wang2013quantifying}
Dashun Wang, Chaoming Song, and Albert-L{\'a}szl{\'o} Barab{\'a}si.
\newblock Quantifying long-term scientific impact.
\newblock {\em Science}, 342(6154):127--132, 2013.

\bibitem{wang2013citation}
Jian Wang.
\newblock Citation time window choice for research impact evaluation.
\newblock {\em Scientometrics}, 94(3):851--872, 2013.

\bibitem{wang2014unpacking}
Jian Wang.
\newblock Unpacking the {M}atthew effect in citations.
\newblock {\em Journal of Informetrics}, 8(2):329--339, 2014.

\bibitem{wang2022self}
Xuezhi Wang, Jason Wei, Dale Schuurmans, Quoc Le, Ed~Chi, Sharan Narang,
  Aakanksha Chowdhery, and Denny Zhou.
\newblock Self-consistency improves chain of thought reasoning in language
  models.
\newblock {\em arXiv preprint arXiv:2203.11171}, 2022.

\bibitem{wei2022chain}
Jason Wei, Xuezhi Wang, Dale Schuurmans, Maarten Bosma, Fei Xia, Ed~Chi, Quoc~V
  Le, Denny Zhou, et~al.
\newblock Chain-of-thought prompting elicits reasoning in large language
  models.
\newblock {\em Advances in neural information processing systems},
  35:24824--24837, 2022.

\bibitem{wu2019large}
Lingfei Wu, Dashun Wang, and James~A Evans.
\newblock Large teams develop and small teams disrupt science and technology.
\newblock {\em Nature}, 566(7744):378--382, 2019.

\bibitem{yao2024tree}
Shunyu Yao, Dian Yu, Jeffrey Zhao, Izhak Shafran, Tom Griffiths, Yuan Cao, and
  Karthik Narasimhan.
\newblock Tree of thoughts: Deliberate problem solving with large language
  models.
\newblock {\em Advances in Neural Information Processing Systems}, 36, 2024.

\bibitem{Fang2024disciplines}
Fang Zhang and Shengli Wu.
\newblock Predicting citation impact of academic papers across research areas
  using multiple models and early citations.
\newblock {\em Scientometrics}, 129(7):4137--4166, 2024.

\bibitem{zheng2025large}
Yizhen Zheng, Huan~Yee Koh, Jiaxin Ju, Anh~TN Nguyen, Lauren~T May, Geoffrey~I
  Webb, and Shirui Pan.
\newblock Large language models for scientific discovery in molecular property
  prediction.
\newblock {\em Nature Machine Intelligence}, pages 1--11, 2025.

\bibitem{ziems2024can}
Caleb Ziems, William Held, Omar Shaikh, Jiaao Chen, Zhehao Zhang, and Diyi
  Yang.
\newblock Can large language models transform computational social science?
\newblock {\em Computational Linguistics}, 50(1):237--291, 2024.

\end{thebibliography}

\clearpage
\appendix

\setcounter{figure}{0}
\renewcommand{\thefigure}{A\arabic{figure}}
\setcounter{table}{0}
\renewcommand{\thetable}{A\arabic{table}}

\section*{Appendix}
\label{sec:appendix}

\subsection*{SciSciNet data}
We use the SciSciNet \citep{lin2023sciscinet} dataset, a large-scale bibliometric database, to construct our reference sample (\url{https://doi.org/10.6084/m9.figshare.c.6076908.v1}). The selected variables and their descriptions are provided in Appendix Table \ref{tab:sciscinet}. Merging all tables yields a total of $133,518,311$ papers. From this corpus, we apply a series of selection criteria, restricting our sample to papers published in Q1 journals between $1999$ and $2021$, containing $3$ to $54$ references, and having received at least one citation. Additionally, we require focal papers to have an assigned ``top field'' classification and a valid DOI with an abstract, resulting in a filtered dataset of $17,538,900$ papers. Abstracts are retrieved via CrossRef and Semantic Scholar APIs \citep{crossref_api,kinney2023semantic}. From this refined dataset, we randomly sample $10,000$ focal papers for our analysis.

\subsection*{Prompting GPT-4o}
We use a straightforward prompting setup for GPT-4o, specifically version \textit{gpt-4o-2024-08-06}. As format restrictions may degrade model performance on some tasks, we do not impose them here \citep{tam2024let}, and instead rely on a dedicated postprocessing step to parse and clean the model outputs. This choice inherently risks slightly degraded performance on certain tasks, but ensures the generation of references remains unconstrained and maximally reflective of the model's parametric knowledge. Our prompting strategy is intentionally simple, though we note that more sophisticated prompting methods \citep{kojima2022zeroshot,fairprompting,wang2022self,wei2022chain,yao2024tree} or deeper reasoning frameworks \citep{guo2025deepseek,muennighoff2025s1simpletesttimescaling} could be investigated to improve existence rates and reduce fabrication of references.

The system prompt for each focal paper is:
\begin{quote}
\textit{Below, we share with you the title, authors, year, venue, and abstract of a scientific paper. Can you provide $\{n\}$ references that would be relevant to this paper?}
\end{quote}

The user prompt contains basic bibliographic metadata and the abstract:
\begin{verbatim}
Title: {PaperTitle}

Authors: {Author_Name}

Year: {Year}

Venue: {Journal_Name}

Abstract: {Abstract}
\end{verbatim}

Here $\{n\}$ is set to the ground-truth number of cited works for that focal paper. We subsequently parse the model's suggestions for bibliometric data (title, authors, number of authors, venue, publication year) in a postprocessing step using a second prompt (called via \textit{gpt-4o-mini-2024-07-18}):

\begin{quote}
\textit{Below, we share with you a list of references. Could you for each reference extract the authors, the number of authors, title, publication year, and publication venue? Please only return the extracted information in a markdown table with the authors, number of authors, title, publication year, and publication venue as columns. Do not return any additional information or formatting.}
\end{quote}

This postprocessing prompt enforces a structured output format while minimizing reliance on stringent formatting instructions during the reference generation step itself.

\subsection*{Existence check}
We assess whether each generated reference has a corresponding record in the SciSciNet \citep{lin2023sciscinet} database using a two-step process: (1) retrieving candidate matches via Elasticsearch \citep{elasticsearch}, and (2) applying fuzzy matching on both titles and authors. We manually verified a random sample of 100 references by cross-checking the titles and authors on Google Scholar. Within these 100 references, our existence algorithm identified 42 as existing, and all were correctly classified. Of the 58 non-existent, five actually exist. This shows that our existence algorithm may underestimate the actual existence rates of LLM-generated references by only using the SciSciNet \citep{lin2023sciscinet} dataset.

\paragraph{Elasticsearch Retrieval.}
We rely on Elasticsearch \citep{elasticsearch} to search the ``Title'' field in the SciSciNet index, which uses a well-known ranking approach called BM25 \citep{schutze2008introduction}. Elasticsearch returns the three best-matching records for every generated reference title. This ensures we capture the most relevant candidates, even if there are small variations in spelling or spacing.

\paragraph{Fuzzy Matching.}
From the three candidates, we compare both the reference's title and its list of authors to each candidate's stored metadata. For the title comparison, we use partial string matching based on the Levenshtein distance, allowing us to handle truncated or rearranged segments \citep{christen2012data}. For the authors, we check for common tokens across the two author strings, again leveraging the Levenshtein distance to accommodate partial matches and differences in order \citep{christen2012data}. Each candidate is assigned two scores—one for title similarity and one for author similarity—which we then average into a single overall match score.

\paragraph{Existence Thresholds.}
We select the candidate with the highest overall match score. To confirm it as a genuine match, the title similarity must exceed a high threshold of $0.90$ (indicating near-identical titles), while the author similarity must surpass a more moderate threshold of $0.50$ (reflecting adequate alignment in author strings). Only if both conditions are met do we label the reference as ``existing.'' Otherwise, it is deemed ``non-existent.''

\subsection*{Pairwise two-sided Wilcoxon signed-rank test}
Comparing distributions of bibliometric data, such as citation or reference counts, is challenging due to skewness, heavy tails, and potential within-paper dependencies. Additionally, because we generate references solely from each focal paper's title and abstract—rather than having a one-to-one correspondence with a paper's actual references—we cannot directly match references between ground truth and GPT-4o outputs. Instead, for each focal paper, we compute the median of the variable of interest (e.g., the median citation count) for its ground truth references and for the (existing or non-existent) LLM-generated references. We then consider the distribution of these per-paper differences across all focal papers in a given field. The Wilcoxon signed-rank test then evaluates whether these differences systematically deviate from zero. This approach avoids strong distributional assumptions or independence requirements, making it well-suited for the heavy-tailed nature of citation data and for the potential dependencies among references within a single focal paper.

\subsection*{Embeddings}
We measure content-level alignment between focal papers and references by comparing the cosine similarity of their textual embeddings. Specifically, for each focal paper (title and abstract) and each reference (title only), we compute embeddings using either OpenAI \texttt{text-embedding-3-large} \citep{arts2023beyond,kusupati2022matryoshka} or SPECTER2 \citep{singh-etal-2023-scirepeval}. The vector representations capture semantic relationships between documents, enabling a pairwise cosine similarity metric that quantifies how semantically aligned a reference is to the focal paper. The OpenAI text-embedding-3-large model generates $3,072$-dimensional embeddings. To confirm the robustness of our results, we repeat the same similarity comparisons using SPECTER2, a $768$-dimensional embedding model designed for scientific texts. 

We experiment with two settings: (1) embeddings computed from titles only, and (2) embeddings computed from both titles and abstracts. For each focal paper, we also compute cosine similarities with references drawn at random from another ground truth paper in the same field. These random reference scores serve as a baseline for context-unaware matching.

\subsection*{Citation networks}
Let $G=(V, \: E)$ be a simple undirected graph, where $V$ denotes the set of nodes and $E$ denotes the set of edges. Let $|V|$ denote the number of nodes. The \textit{distance} between two nodes $u, \: v \in V$ is denoted by $d(u, v)$ and equals the numbers of edges in the shortest path between the two nodes. The \textit{average shortest path length} $L$ of the graph $G$ is given by
\begin{equation*}
L = \frac{1}{|V|(|V|-1)} \sum_{u \neq v \in V} d(u,v) \,.
\end{equation*}
\textit{The density} $D$ of $G$ is the number of edges of the graph relative to the number of possible edges between all the nodes in the graph
\begin{equation*}
    D = \frac{|E|}{\binom{|V|}{2}} = \frac{2|E|}{|V|(|V|-1)} \,.
\end{equation*}
For a node $v \in V$ with degree $d_v$, let $e(v)$ denote the number of edges among its neighbors. The \textit{clustering coefficient} of $v$ is defined as 
\begin{equation*}
    C(v) = \frac{2\,e(v)}{d_v\,\bigl(d_v - 1\bigr)} \,.
\end{equation*}
The \textit{average clustering coefficient} $\mu_{C}$ of $G$ is then
\begin{equation*}
    \mu_{C} = \frac{1}{|V|} \sum_{v \in V} C(v) \,.
\end{equation*}
For a node $v \in V$ with degree $d(v)$, the \textit{degree centrality} $C_D(v)$ can be defined as
\begin{equation*}
    C_D(v) = \frac{d_v}{|V| - 1} \,,
\end{equation*}
reflecting the fraction of possible connections that $v$ actually has. The \textit{closeness centrality} of a node $v \in V$ is determined by the reciprocal of the average shortest path from $v$ to all other nodes:
\begin{equation*}
    C_C(v) = \frac{|V|-1}{\sum_{u \neq v} d(u, v)} \,.
\end{equation*}
Let $C_E(v)$ be the \textit{eigenvector centrality} of node $v \in V$ defined as
\begin{equation*}
    C_E(v) = \frac{1}{\lambda}\sum_{u \in \mathcal{N}(v)}C_E(u) \,,
\end{equation*} 
where $\lambda$ is the largest eigenvalue of the adjacency matrix of the graph. The \textit{standard deviation of the eigenvector centrality}, \(\sigma_{C_E}\), is
\begin{equation*}
    \sigma_{C_E} = \sqrt{\frac{1}{|V|}\sum_{v \in V} \Bigl(C_E(v) - \mu_{C_E}\Bigr)^2} \,,
\end{equation*}
where $\mu_{C_E}$ is the mean of the eigenvector centralities.

\clearpage

\section*{Figures}
\label{sec:figures}

\vfill

\begin{figure}[h!]
	\centering
	\includegraphics[width=1\textwidth]{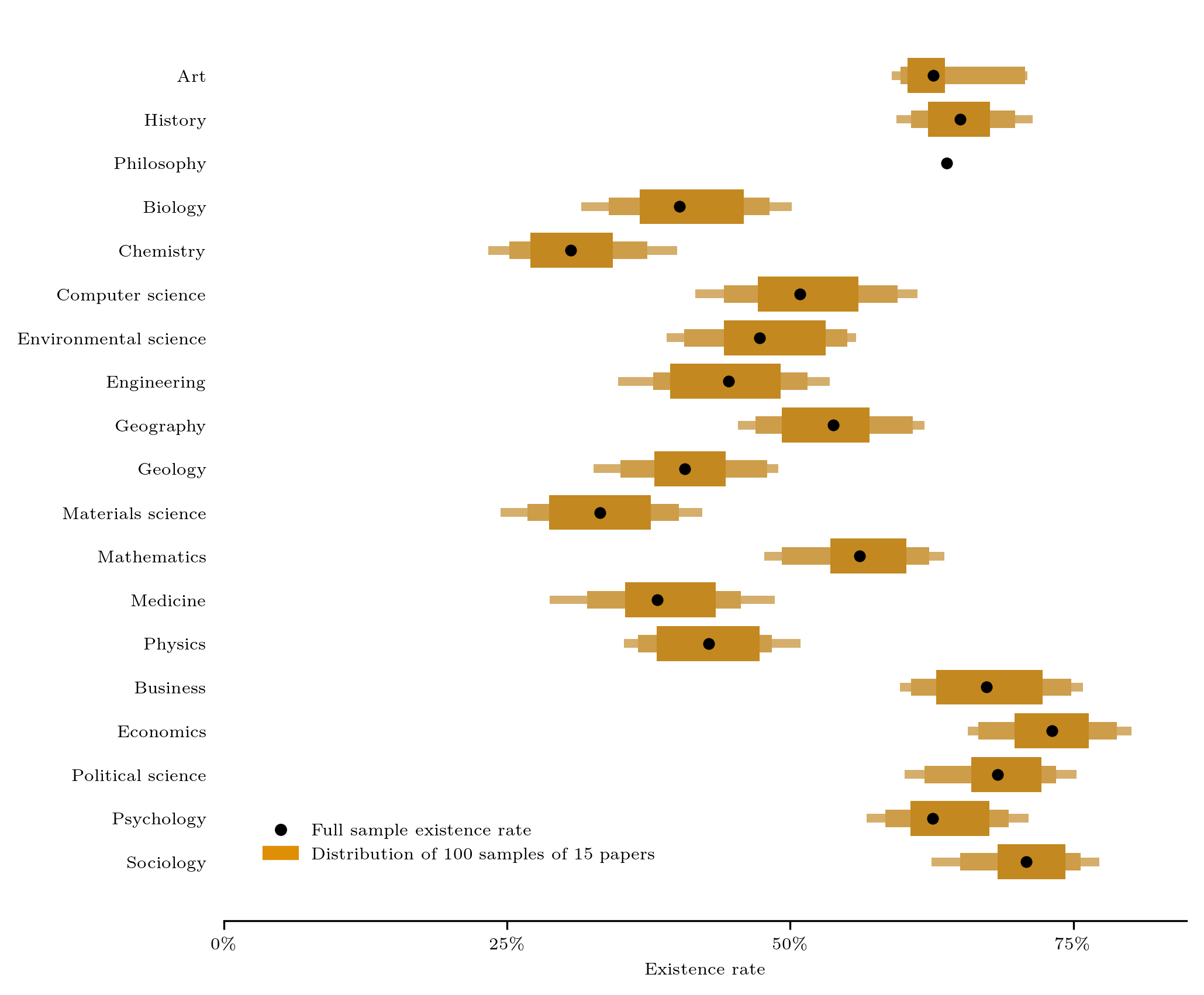}
	\caption{\textbf{The field-level variation in existence rates persists even when controlling for the varying sample sizes of focal papers across fields.} This figure displays the distributions (n=$100$) of the existence rates of generated references across fields when we subsample $15$ random focal papers from the same field. The subsample size of $15$ corresponds to the number of focal papers in the field of philosophy which has the lowest number of focal papers.}
	\label{fig:appendix_3}
\end{figure}

\vfill

\begin{figure}[t!]
	\centering
	\includegraphics[width=1\textwidth]{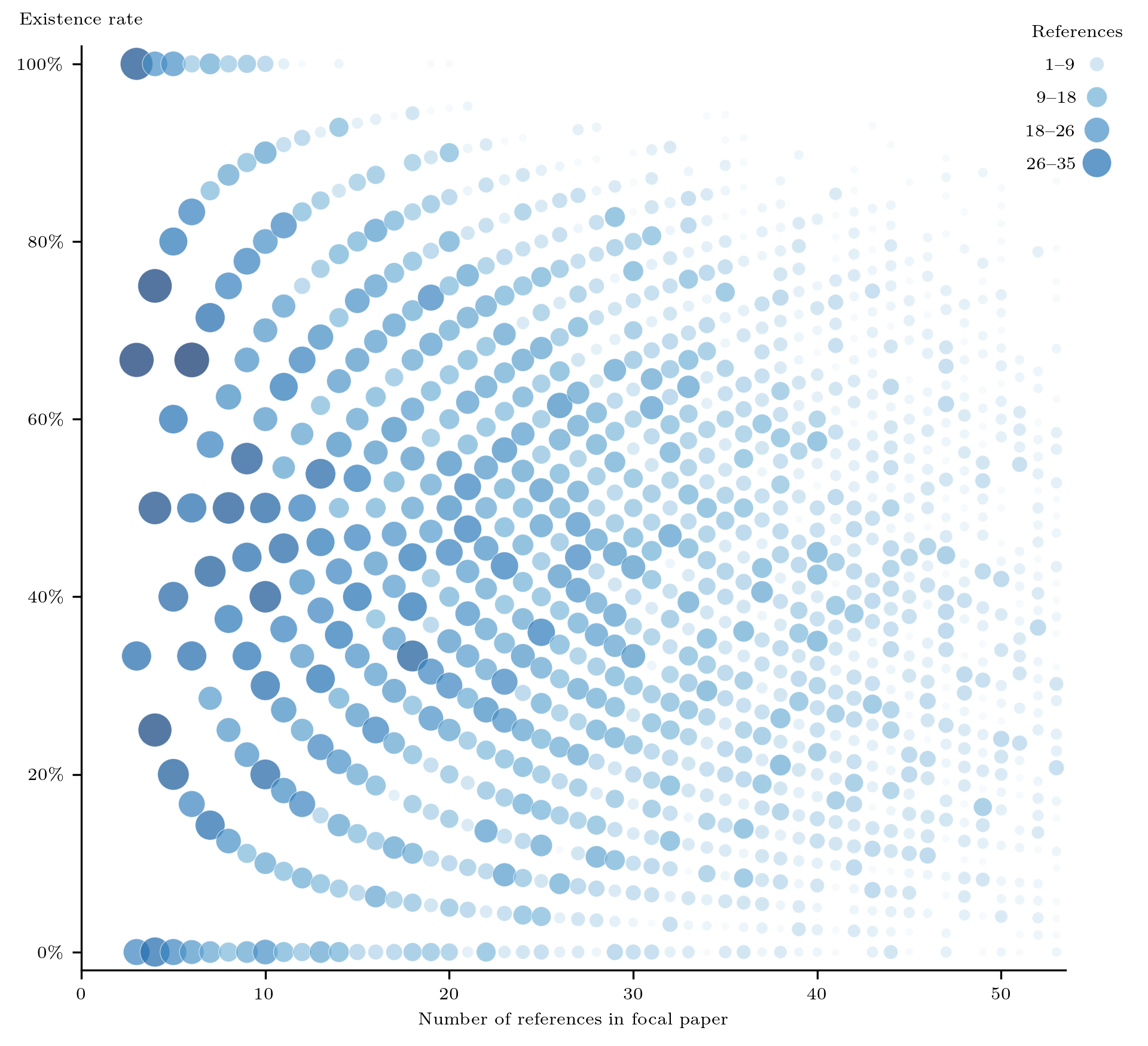}
	\caption{\textbf{The existence rates on the focal paper level show only moderate negative correlation with the number of requested references.} This figure displays the existence rates on the focal paper level (n=$10,000$) for a varying number of references. The dot size represents the total number of focal papers for a given existence rate and number of references. We see only a moderate negative correlation ($-0.10$, $p$$<$$0.001$), indicating that a large number of references is not the main driver behind a lower existence rate.}
	\label{fig:appendix_2}
\end{figure}

\begin{figure}[t!]
	\centering
	\includegraphics[width=1\textwidth]{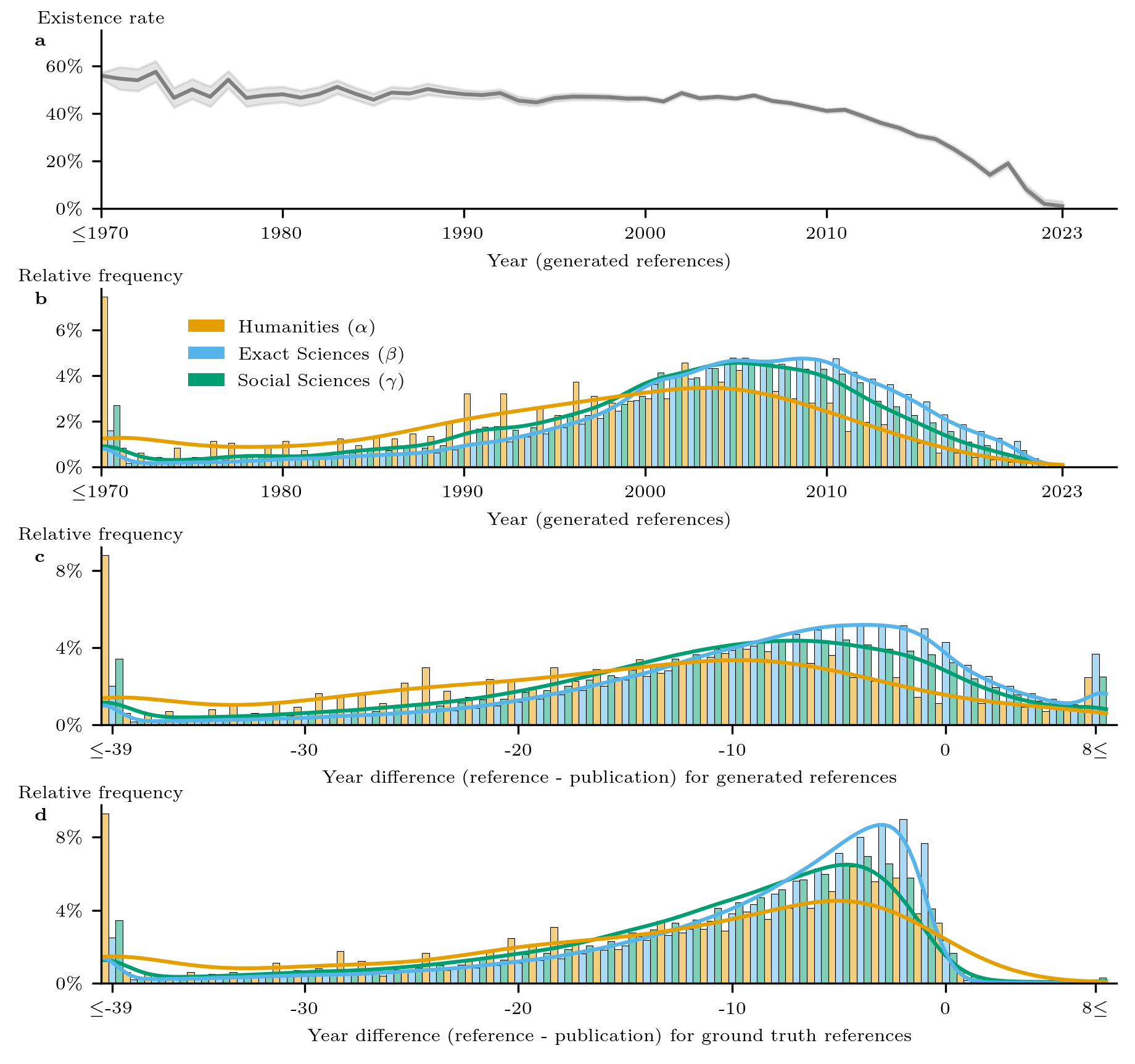}
	\caption{\textbf{The higher existence rates in humanities and social sciences can be partially attributed to their tendency to cite older references, both in absolute terms and relative to the publication year of the focal paper.} This figure displays the existence rate of generated references (n=$274,497$) over time, the absolute distribution of reference publication years for generated references, and the relative distribution of reference publication years for generated and ground truth references (n=$274,951$) with respect to the focal paper's publication year. Shaded bands represent $95\%$ confidence intervals. \textbf{a}, The existence rate of generated references as a function of their publication year, with a declining trend for more recent references. \textbf{b}, The absolute distribution of reference publication years for three different categories ($\alpha$, $\beta$, $\gamma$, see Appendix Table \ref{tab:mapping}), showing the overall temporal patterns in generated references. \textbf{c}, The relative distribution of reference publication years for generated references, measured as the difference between the reference's publication year and that of the focal paper. \textbf{d}, The relative distribution of reference publication years for ground truth references, serving as a comparison to panel (c) and showing how human citation practices distribute references over time.}
	\label{fig:appendix_4}
\end{figure}

\begin{figure}[t!]
	\centering
	\includegraphics[width=1\textwidth]{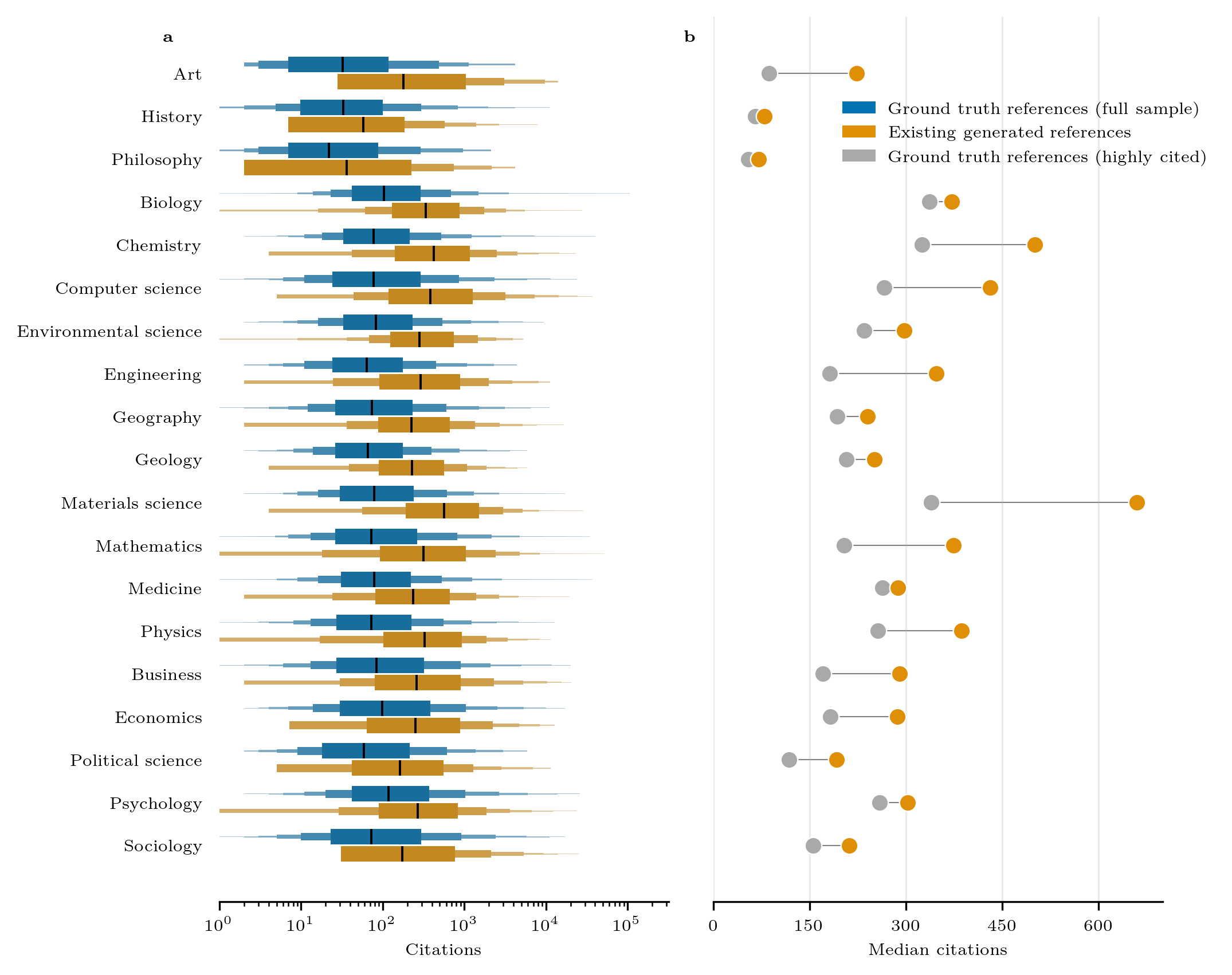}
	\caption{\textbf{Higher citation counts in existing generated references are not merely due to selection bias.} This figure displays the distribution of citations for generated (orange, n=$116,939$) and ground truth references (blue, n=$274,951$), and the median citation counts for generated references (orange, n=$116,939$) compared to the most highly cited ground truth references (gray, n=$116,939$), selected by matching, for each focal paper, the number of existing generated references. \textbf{a}, The distribution of citations for generated and ground truth references shows there is only small overlap. \textbf{b}, The observed gap in median citation counts between existing generated references and ground truth references remains even when selecting, for each focal paper, the same number of most highly cited ground truth references as the number of existing generated references. This confirms that the higher citation counts in existing generated references are not merely due to selection bias. The pairwise two-sided Wilcoxon signed-rank test at the focal paper level confirms that the existing generated references gap remains statistically significant for all fields ($p$$<$$0.05$), except history, philosophy, geography, geology, and medicine.}
	\label{fig:appendix_11}
\end{figure}

\begin{figure}[t!]
	\centering
	\includegraphics[width=1\textwidth]{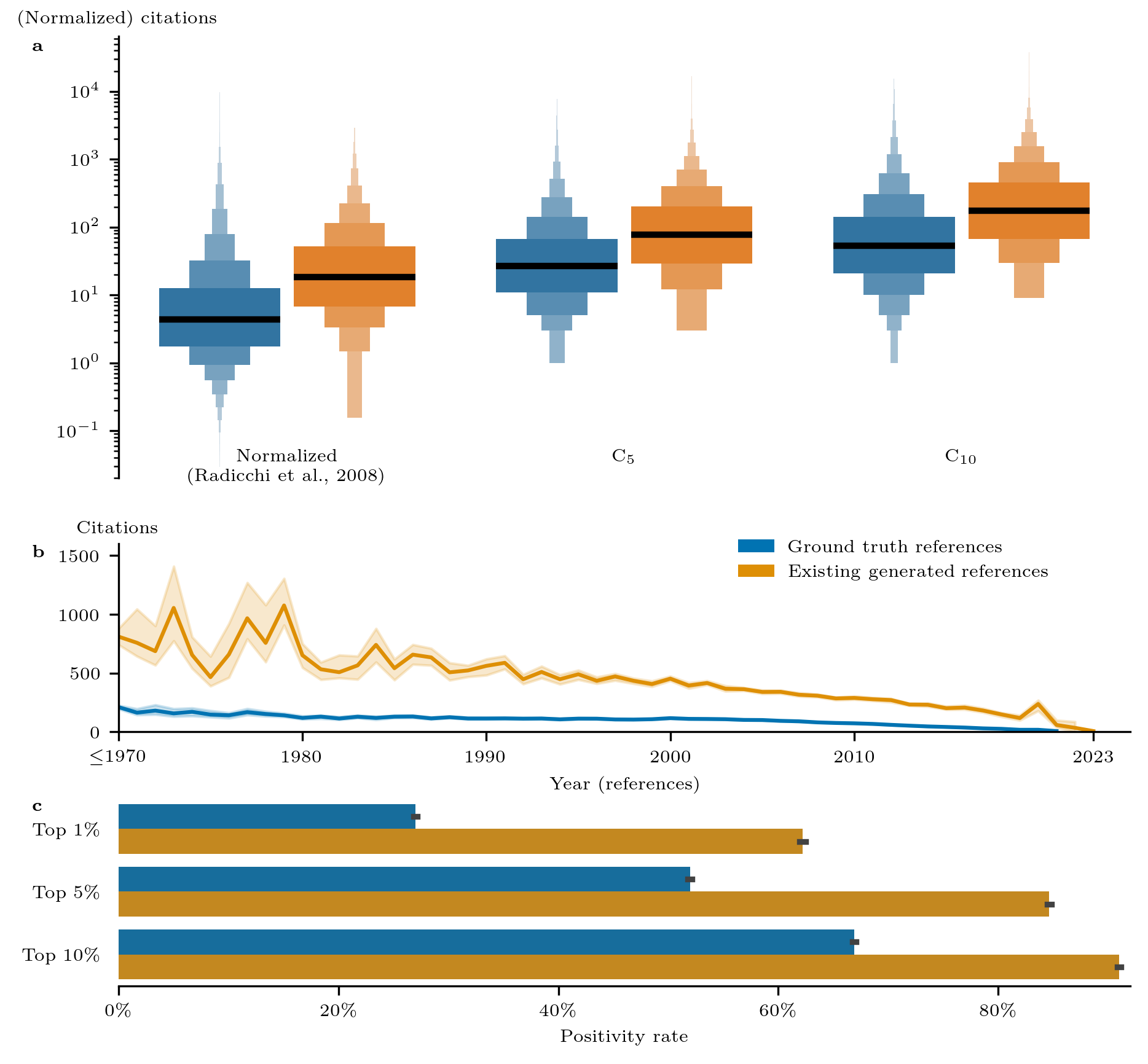}
	\caption{\textbf{Existing generated references exhibit consistently higher citation counts across multiple citation impact metrics.}  This figure presents alternative citation impact metrics comparing ground truth references (blue, n=$274,951$) and existing generated references (orange, n=$116,939$). Error bars and shaded bands represent $95\%$ confidence intervals. \textbf{a}, The distribution of (normalized) citation counts across different citation metrics: normalized citations \citep{radicchi2008universality}, total citations after 5 years (C$_5$), and total citations after 10 years (C$_{10}$). Existing generated references consistently show higher median citation counts across all metrics. \textbf{b}, The median citation count of ground truth and existing generated references as a function of their publication year. The consistently higher citation counts for existing generated references suggest that their impact is not simply due to a recency bias but rather a systematic preference for highly cited works. \textbf{c}, The proportion of ground truth and existing generated references appearing in the top $10\%$, $5\%$, and $1\%$ most cited papers within their respective field and publication year \citep{lin2023sciscinet}. Approximately $90\%$ of existing generated references fall within the top $10\%$, and over $60\%$ appear in the top $1\%$, more than twice the rate observed for ground truth references, reinforcing the Matthew effect in citations.}
	\label{fig:appendix_6}
\end{figure}

\begin{figure}[t!]
	\centering
	\includegraphics[width=1\textwidth]{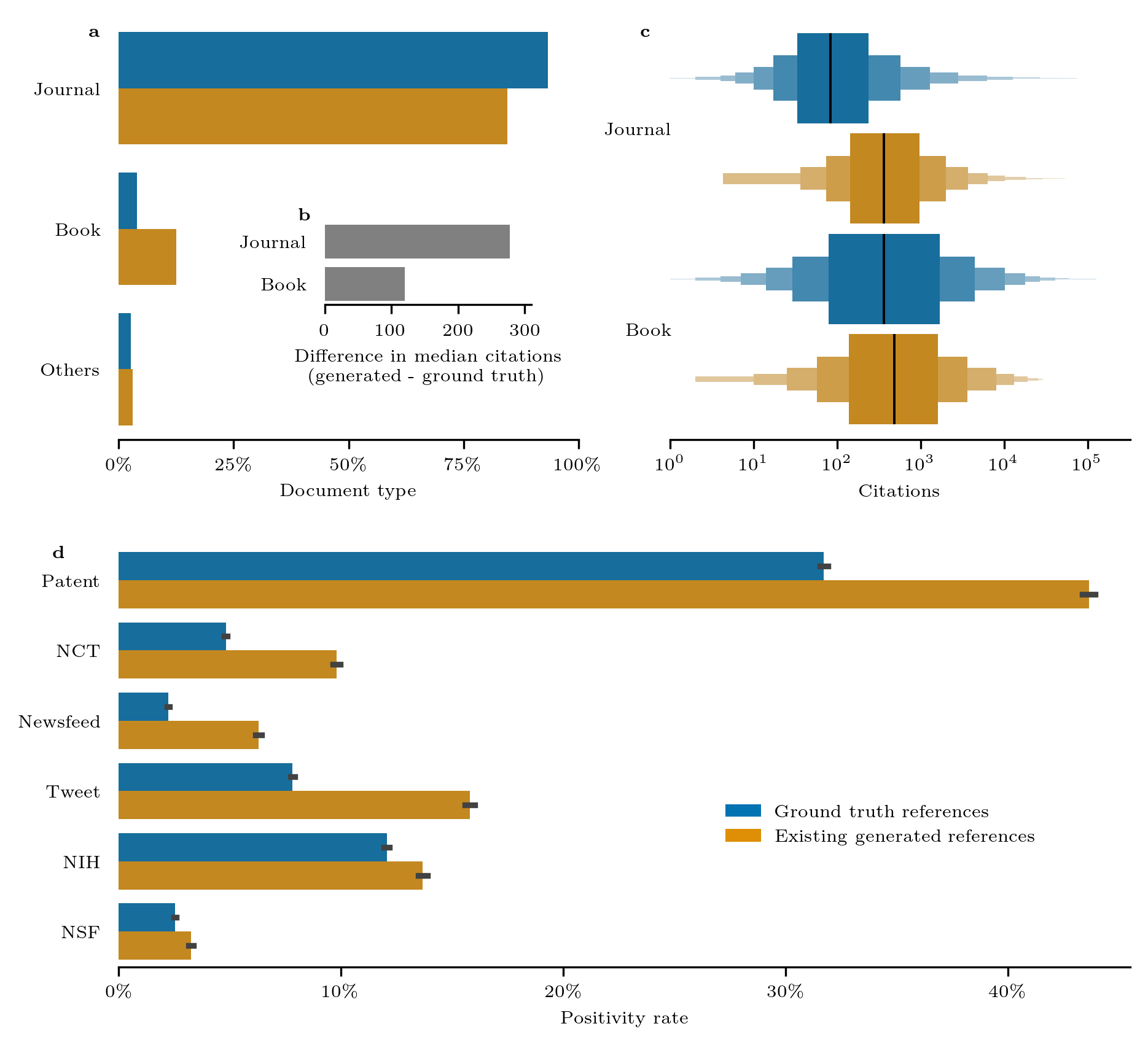}
	\caption{\textbf{Differences in document types, citation impact, and broader influence between ground truth and existing generated references.} This figure compares the distribution of document types, citation counts, and alternative impact indicators between ground truth references (blue, n=$274,951$) and existing generated references (orange, n=$116,939$). Error bars represent $95\%$ confidence intervals. \textbf{a}, The distribution of document types, showing that both ground truth and generated references are predominantly journal articles, though existing generated references contain a slightly higher proportion of books. \textbf{b}, The difference in median citation counts for journals and books, highlighting that generated references tend to have higher citation counts, even when comparing journals and books separately. \textbf{c}, The citation distribution for journal articles and books, illustrating that existing generated references, even when limited to journal articles, achieve citation counts comparable to books in the ground truth references. \textbf{d}, The prevalence of alternative impact indicators, including citations in patents, clinical trials (NCT), news articles, tweets, and funding acknowledgments from NIH and NSF grants. Existing generated references exhibit consistently higher rates across these alternative impact measures, indicating a broader influence beyond academic citations.}
	\label{fig:appendix_5}
\end{figure}

\begin{figure}[t!]
	\centering
	\includegraphics[width=1\textwidth]{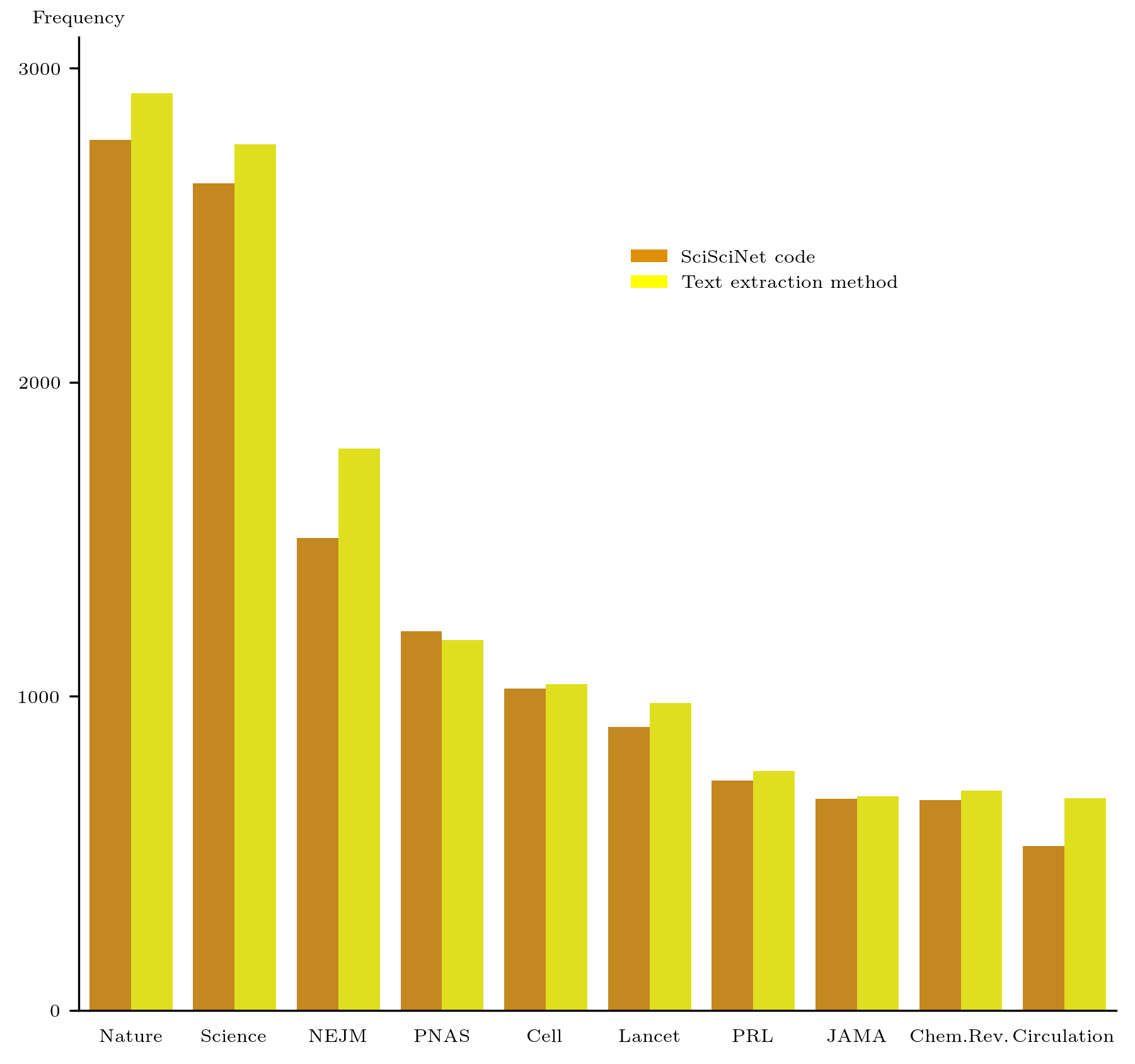}
	\caption{\textbf{Comparison of journal name identification methods for ground truth and generated references.} This figure compares the frequency of top journal venues identified using two different methods: the SciSciNet \citep{lin2023sciscinet} database (brown) and a validated text extraction method (yellow). For ground truth and existing generated references, publication venues are directly obtained from the SciSciNet \citep{lin2023sciscinet} database. For non-existent generated references, where no direct database match is available, we extract journal names using exact matches for the official journal names and ISO-4 abbreviations. The comparison across major journals such as Nature, Science, NEJM, PNAS, Cell, and Lancet shows that the text extraction method produces similar frequency distributions to SciSciNet \citep{lin2023sciscinet} for the existing generated references, validating its reliability for identifying publication venues.}
	\label{fig:appendix_1}
\end{figure}

\begin{figure}[t!]
	\centering
	\includegraphics[width=1\textwidth]{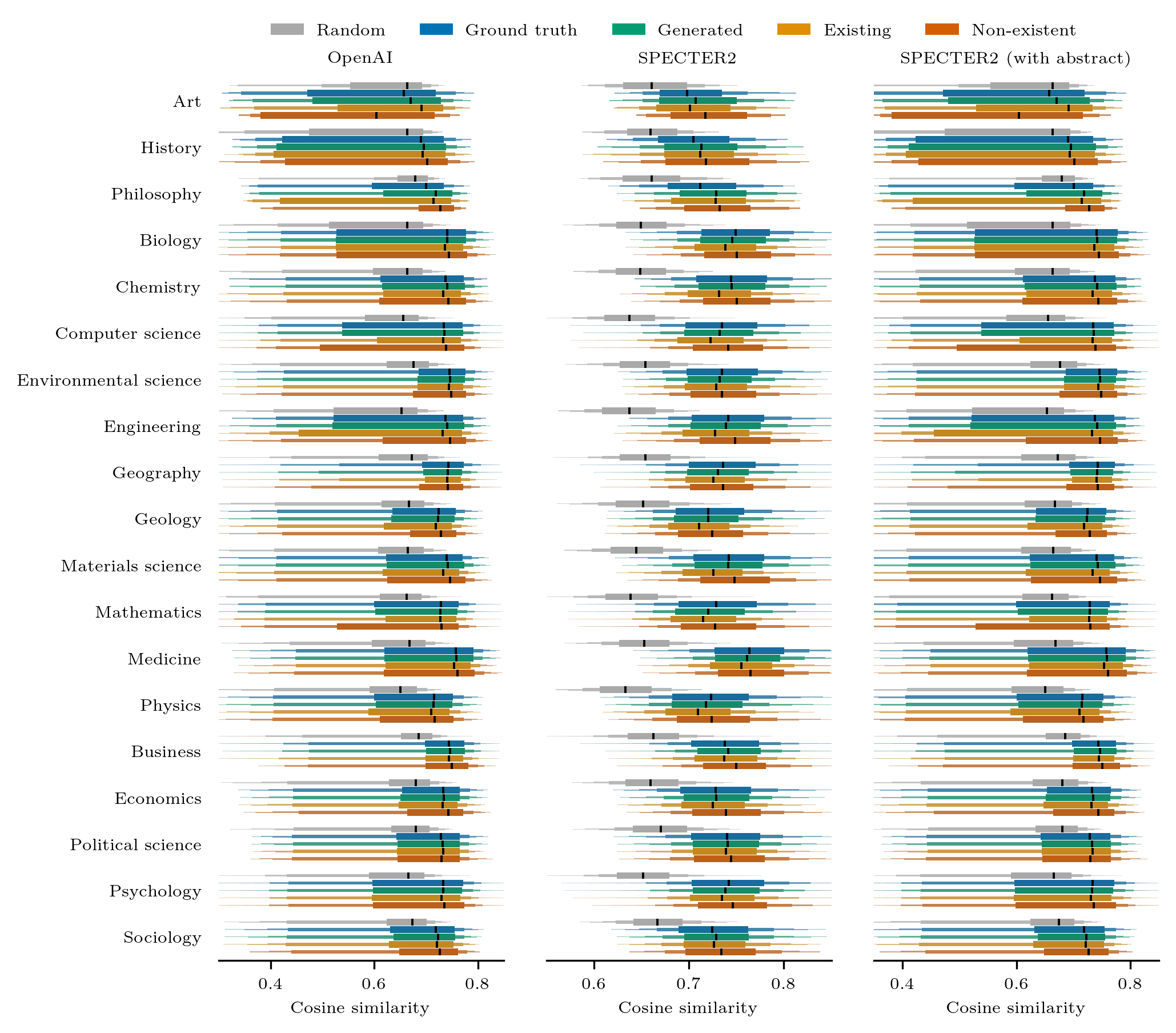}
	\caption{\textbf{Cosine similarity between focal papers and references using different embedding models and input configurations.} This figure displays the distributions of pairwise cosine similarity scores between focal papers and their references across scientific fields, using different embedding models and levels of textual input. The left panel shows cosine similarity computed using OpenAI text-embedding-3-large (size$=3,072$) applied only to the titles of focal papers. The middle panel presents results using SPECTER2 embeddings (size$=768$) applied only to titles, while the right panel extends SPECTER2 embeddings to both titles and abstracts of the focal papers. The similarity distributions are shown for ground truth references (blue, n=$274,951$), all generated references (green, n=$274,497$), existing generated references (orange, n=$116,939$), and non-existing generated references (red, n=$157,558$). A benchmark comparison is provided using a random ground truth reference list from the same field (gray, n=$274,951$). Across all embedding methods, generated references exhibit cosine similarity scores comparable to ground truth references and consistently higher than random references.}
	\label{fig:appendix_8}
\end{figure}

\begin{figure}[t!]
	\centering
	\includegraphics[width=1\textwidth]{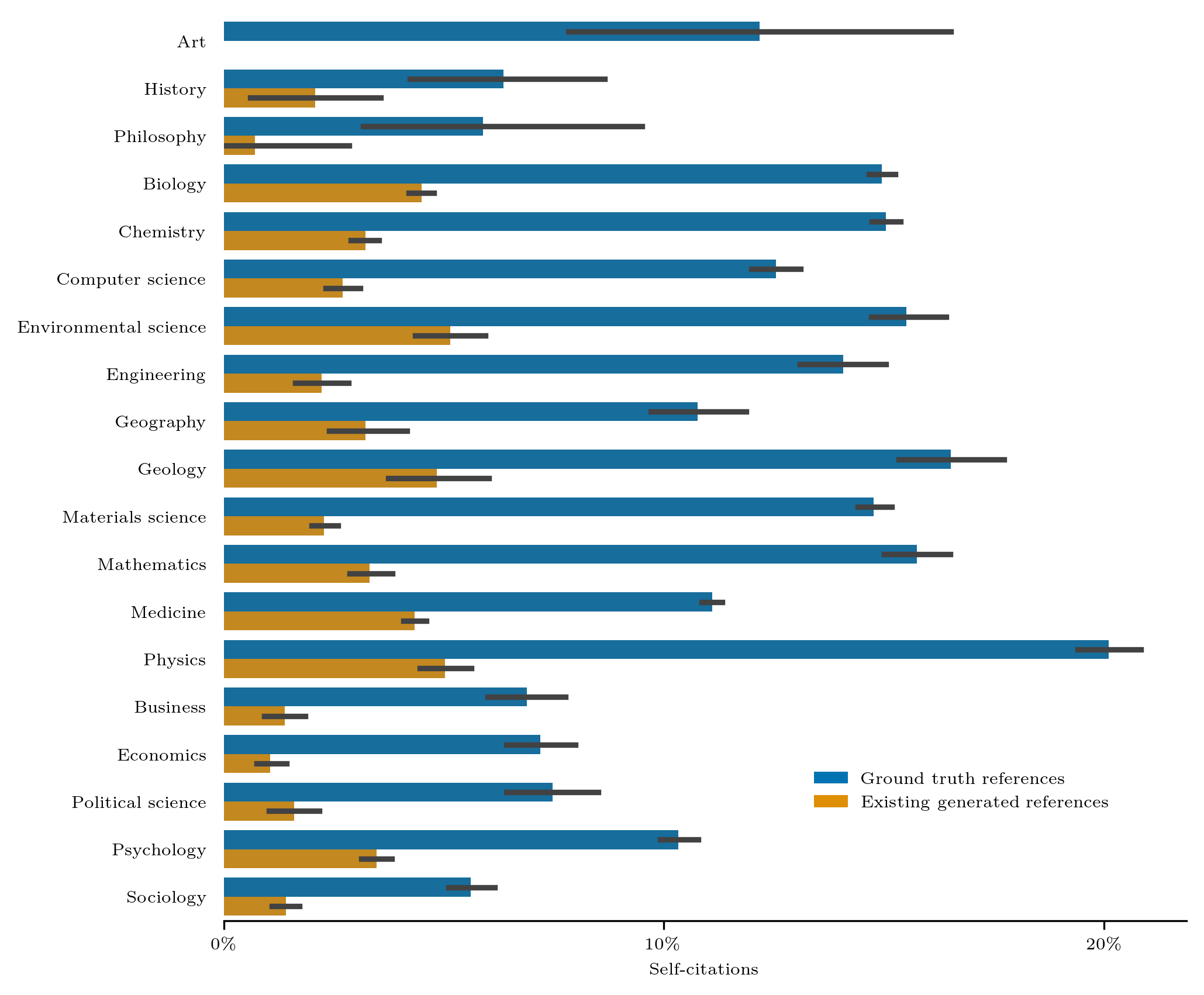}
	\caption{\textbf{LLM-generated references exhibit consistently lower self-citation rates across scientific fields compared to ground truth references.} This figure presents the proportion of author-level self-citations for ground truth references (blue, n=$274,951$) and existing generated references (orange, n=$116,939$) across different scientific disciplines. Self-citations are defined as instances where at least one author of the focal paper cites another work that includes their name, identified using SciSciNet \citep{lin2023sciscinet} author-matching codes. While the self-citation rates of existing generated references follow a similar field-specific variation as ground truth references \citep{Szomszor2020self}, they are consistently lower across all disciplines. This suggests that LLMs do not replicate human self-referential citation patterns at the same rate, potentially reducing self-citation biases and altering the dynamics of author visibility and impact. Error bars represent $95\%$ confidence intervals.}
	\label{fig:appendix_7}
\end{figure}

\begin{figure}[t!]
	\centering
	\includegraphics[width=1\textwidth]{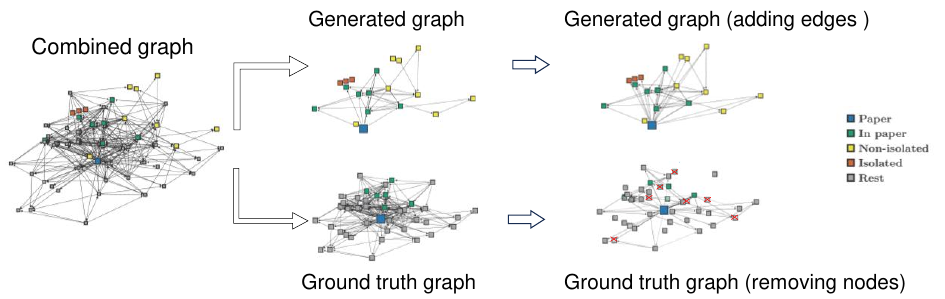}
	\caption{\textbf{The pipeline depicting the generation of the local citation graphs corresponding to the ground truth references and the LLM-generated references.} This figure displays the local citation graph of the focal paper consisting of both the ground truth and LLM-generated references. We then show how the full graph is split up into two distinct citation graphs used for the analysis in Figure \ref{fig:main_6}. Here, we distinguish between references that are both cited by the focal paper (blue node) and suggested by GPT-4o (green nodes), non-isolated generated references that that are not cited by the focal paper but still connected to at least one other reference (ground truth or generated) (yellow nodes), isolated generated references that are not cited by the focal paper and also not connected to any other reference (orange nodes), and the remaining ground truth references that are not suggested by GPT4-o (grey). An arrow between $A$ and $B$ corresponds to a citation from $A$ to $B$. We use the SciSciNet \citep{lin2023sciscinet} dataset to retrieve this citation information. For each of the $10,000$ focal papers (blue), we construct the full graph consisting of both the ground truth references and the existing references suggested by GPT4-o as displayed on the left hand side. The full graph is then split up into the sub-graph corresponding to the focal paper and the LLM-generated references (blue, green, yellow and orange nodes) and the sub-graph corresponding to the focal paper and the ground truth references (blue, green and grey nodes). To ensure a fair comparison, each of those two graphs undergoes a further post-processing step. First, for the generated graph, we add an edge from the focal paper to the generated references not initially cited by the focal paper (yellow and orange nodes) to ensure the connectivity of the graph. Second, for ground truth graph, since approximately $50\%$ of the GPT4-o-generated references exist (see Fig. \ref{fig:main_3}\textbf{a}), we randomly remove a subset of references (green and grey nodes) to ensure it is equal in size to the generated graph on a focal paper level. For analytical simplicity both graphs were converted to undirected graphs.}
	\label{fig:appendix_9}
\end{figure}

\begin{figure}[t!]
	\centering
	\includegraphics[width=1\textwidth]{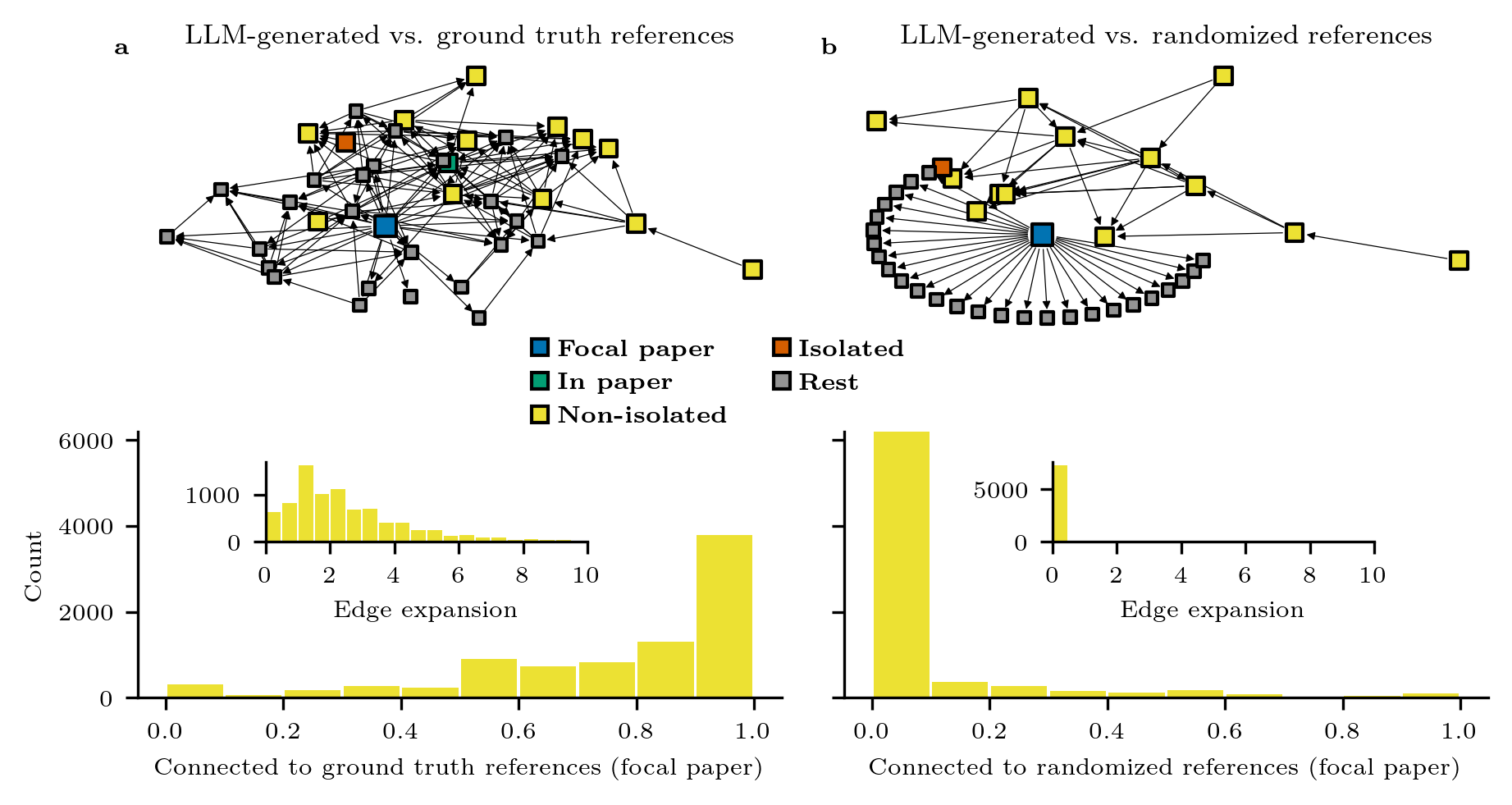}
	\caption{\textbf{LLM-generated references are connected more closely to the ground truth references compared to the randomized references.} This figure displays that the references generated by GPT4-o are well connected to the ground truth references despite not being cited by the focal paper itself. A similar connectivity is not observed for the randomized ground truth references. \textbf{a,} In the top, we display the local citation context of the ground truth and LLM-generated references of a focal paper (blue node). Here, we distinguish between references that are both cited by the focal paper and suggested by GPT-4o (green nodes), non-isolated generated references that that are not cited by the focal paper but still connected to at least one other reference (ground truth or generated) (yellow nodes), isolated generated references that are not cited by the focal paper and also not connected to any other reference (orange nodes), and the remaining ground truth references that are not suggested by GPT4-o (grey). An arrow between $A$ and $B$ corresponds to a citation from $A$ to $B$. As suggested by the depicted graph, among the LLM-generated references and per focal paper, the majority of them are non-isolated (mean: $54\%$, median: $60 \%$, standard deviation: $28 \%$), whereas less are isolated (mean: $37\%$, median: $30 \%$, standard deviation: $30 \%$) or appear in the paper (mean: $9\%$, median: $4 \%$, standard deviation: $13 \%$). Below, we then display the fraction of non-isolated, LLM-generated references (per focal paper) that are connected to at least one ground truth reference. Here, we observe that most generated references are indeed citing or cited by at least ground truth references instead of merely being connected to another generated references. For example, for the graph displayed above this number is equal to $92 \%$. This connectivity is further reflected in the edge expansion---the number of edges between ground truth references and non-isolated GPT-4o-generated references, normalized by the number of non-isolated GPT-4o-generated references---showing strong overall linkage (e.g., $3.58$ in the depicted graph). \textbf{b,} The equivalent analyses where we randomly reshuffle the ground truth references while fixing the field of study of the focal paper. Among the LLM-generated references and per focal paper, we now find that the majority of them are either non-isolated (mean: $49\%$, median: $56 \%$, standard deviation: $33 \%$) or isolated (mean: $51\%$, median: $44 \%$, standard deviation: $33 \%$), whereas a negligible fraction appears in the randomized ground truth references (mean: $0.04\%$, median: $0 \%$, standard deviation: $0.6 \%$). Moreover, among the non-isolated references, most connect only to each other rather than to the randomized references, leading to a fraction of $0\%$ in the example graph. The edge expansion measure is also $0\%$ in that example.  This strong deviation from the observation in (a) suggests that LLMs reflect the local citation context of the focal papers rather well.}
	\label{fig:appendix_10}
\end{figure}

\clearpage 

\section*{Tables}
\label{sec:tables}

\vfill

\renewcommand{\arraystretch}{1.15}

\begin{table}[h!]
\captionsetup{skip=10pt}
\centering
\begin{tabular}{ll}
\toprule
 & \textbf{Field} \\
\midrule
\textbf{Alpha} ($\alpha$) - \textbf{Humanities} & Art \\
                     & History \\
                       & Philosophy \\
\midrule
\textbf{Beta} ($\beta$) - \textbf{Exact sciences} & Biology \\
                      & Chemistry \\
                       & Computer science \\
                       & Environmental science \\
                       & Engineering \\
                       & Geography \\
                       & Geology \\
                       & Materials science \\
                       & Mathematics \\
                       & Medicine \\
                       & Physics \\
\midrule
\textbf{Gamma} ($\gamma$) - \textbf{Social sciences} & Business \\
                      & Economics \\
                       & Political science \\
                       & Psychology \\
                       & Sociology \\
\bottomrule
\end{tabular}
\caption{The categorization of knowledge into distinct fields is based on the framework proposed by \citep{machlup1980knowledge}.}
\label{tab:mapping}
\end{table}

\vfill

\clearpage

\begin{table}[h!]
\captionsetup{skip=10pt}
\centering
\begin{tabular}{lll}
\toprule
& \textbf{Name} & \textbf{Short Description} \\
\midrule
\textbf{Paper level} & & \\
& PaperID & MAG Paper ID of the document. \\
& PaperTitle & Title of the document. \\
& DOI & Digital Object Identifier (DOI) of the document. \\
& DocType & Book, BookChapter, Conference, Dataset, Journal, Repository, \\
&& Thesis, or NULL (unknown). \\
& Year & Publication year of the document. \\
& Reference\_Count & Total reference count of the document. \\
& Citation\_Count & Total citation count of the document. \\
& C5 & The number of citations 5 years after publication. \\
& C10 & The number of citations 10 years after publication. \\
& Patent\_Count & The number of citations by patents from USPTO and EPO. \\
& NCT\_Count & The number of citations by clinical trials from ClinicalTrials.gov. \\
& Newsfeed\_Count & The number of mentions by news from Newsfeed. \\
& Tweet\_Count & The number of mentions by tweets from Twitter. \\
& NIH\_Count & The number of supporting grants from NIH. \\
& NSF\_Count & The number of supporting grants from NSF. \\
\hline
\textbf{Author level} & & \\
& AuthorID & MAG Author ID of the author. \\
& Author\_Name & Original name of the author. \\
\hline
\textbf{Journal level} & & \\
& JournalID & MAG Journal ID of the journal. \\
& Journal\_Name & Original name of the journal. \\
\hline
\textbf{Conference level} & & \\
& ConferenceSeriesID & MAG ConferenceSeries ID of the conference series. \\
& Abbr\_Name & Abbreviated name of the conference series. \\
& ConferenceSeries\_Name & Original name of the conference series. \\
\hline
\textbf{Field level} & & \\
& FieldID & MAG Field ID in the document-field linkage record. \\
& Field\_Name & Original field name of the field of study. \\
& Field\_Type & Top or Sub. Top indicates the top-level field. \\
&& Sub indicates the subfield. \\
& Hit\_1pct & 1 is hit document with top 1\% total citations within the same \\
&& level field and the same year, and 0 is not. \\
& Hit\_5pct & 1 is hit document with top 5\% total citations within the same \\
&& level field and the same year, and 0 is not. \\
& Hit\_10pct & 1 is hit document with top 10\% total citations within the same \\
&& level field and the same year, and 0 is not. \\
& C\_f & Normalized citation as defined by \citep{radicchi2008universality}. \\
\bottomrule
\end{tabular}
\caption{The selection and description of the variables from the SciSciNet \citep{lin2023sciscinet} database.}
\label{tab:sciscinet}
\end{table}

\end{document}